# Space and Time as Containers of the "Physical Material World" with some Conceptual and Epistemological Consequences in Modern Physics

Mauricio Mondragon[*] & Luis Lopez[a]

[a]*Instituto Tecnológico y de Estudios Superiores de Occidente, Departamento de Matemáticas y Física, Periférico Manuel Gómez Morín 8585, C. P. 45090, Tlaquepaque, Jalisco, Mexico*

## Abstract

A particular science is not only defined by its object of study, but also by the point of view and method under which it considers that same object. Taking space and time as an illustrative example, our main aim here is to bring out an almost forgotten conception of science found in many doctrines that seek mainly—but not only—a qualitative and synthetic knowledge rather than, as in modern physics, for example, a quantitative and analytic knowledge. The latter point of view is found to be very limited and fragmented, leaving outside many scientific questions and answers, while the former opens up the way to valuable and interesting answers to those and many other questions. In particular, we argue that the conception of space and time as containers of, respectively, bodies and events clarify many conceptual and epistemological issues of modern science related to the physical material world.

## I.    Introduction

About 60 years ago, Erwin Schrödinger said that the "...image of material reality is today more unsettled and uncertain than it has been for a long time. We know a great many interesting details; every week we learn new ones. However, to pick up from the basic concepts those that have been established as fact and to construct from them a clear and easily understood framework of which we could say: this is certainly so, this we all believe today, is impossible. There is a wide spread hypothesis that an objective image of reality in any previously believed interpretation cannot exist. Only the optimists among us (and I consider myself one of them) consider this a philosophical eccentricity, a desperate measure in the face of a great crisis. We hope that the vacillation of concepts and opinions signifies only an intense process of transformation, which will finally lead to something better than the confused series of formulas that today surround our subject (Heisenberg *et al.* 1961, 46)."

This tendency seems to go on. The confusion and the crisis to which Schrödinger referred at that time have continued. A great majority of modern scientists spend most of their efforts on following the empirical-inductive method of modern science, trying to formulate more or less general theories from specific experimental facts or finding applications to such theories, thus looking away from true principles and completely neglecting the deductive reasoning and the philosophical aspects depending thereon. The philosophical aspects of scientific theories are therefore very largely neglected, at least in the sense herein developed.

---

[*] Corresponding autohor: <momondragon@gmail.com>. Work developed partly during his PhD at *Centre de Physique Théorique* in *Université de la Méditerranée*, Marseilles, France and improved during the following years up to now.

Moreover, in our opinion, this tendency has reached a worrying level, especially when in addition one observes the excessive development of the quantitative aspect of science. Indeed, the present scientific point of view is mainly characterized by seeking to bring everything down to quantity; anything that cannot be so treated is left out of account and is regarded as more or less non-existent or inaccessible. Currently, people commonly think and say that anything that cannot be indicated or represented by numerals, or, in other words, cannot be expressed in purely quantitative terms, lacks scientific value for that reason; and this assumption predominates not only in physics, but in almost all modern science. Even the psychological domain, with the partisans of psycho-physiology, is not beyond its reach.[1] As a consequence, this point of view leaves aside most of the qualitative aspects of the object under study, thus moving away from its nature or essence (in its "Platonic-Aristotelian" sense; see, *e.g.,* Kaminsky 1957), and science splinters into always more specialized fields that are more and more difficult to put together; let alone to find a common objective toward which all of them tend or a common principle synthesizing them.

We believe this *de facto* situation to be very unfortunate, since it strongly disregards and jeopardizes the cognitive value of science, which should be its *raison d'être*. In our opinion, science should try, as much as possible, not to neglect the qualitative aspects of things, which express, better than a "mere" quantitative description, the nature of the object under study. Almost unavoidably, quantitative description implies the extremely limited conception of a physical object as a conglomerate of measurable properties (a striking example is given by the "conventional" formulation of quantum mechanics). In this sense, our general purpose here is to bring back and treat concisely some qualitative aspects of space and time, as well as the role these entities play in the "physical material world", or "corporeal world."[2] In particular, our main concern is to present the idea of space and time as containers of, respectively, bodies and events. The divisibility of space and the unrepeatability of events will also be studied to some extent. Along with these considerations, some conceptual, philosophical, as well as epistemological aspects of modern theories of physics (and science in general) are contrasted with the ideas set out here.

To this purpose, we believe that the best approach is to refer to those who, no matter when or where, have the most comprehensive conception of both science and the universe and, whenever possible, to link their ideas to the modern scientific findings (our considerations embrace mainly their conceptual and epistemological aspects). In this sense, we found Hindu doctrine to be particularly rich and powerful, but it is not the only one. Some Greek "schools" (especially the Pre-Socratics), Far-Eastern doctrines, and the Scholastics of

---

[1] This tendency was already quite definite in the nineteenth century: "In physical science the first essential step in the direction of learning any subject is to find principles of numerical reckoning and practicable methods for measuring some quality connected with it. I often say that when you can measure what you are speaking about, and express it in numbers, you know something about it; but when you cannot measure it, when you cannot express it in numbers, your knowledge is of a meagre and unsatisfactory kind; it may be the beginning of knowledge, but you have scarcely in your thoughts advanced to the state of Science, whatever the matter may be (Kelvin 1891, 80-I)." In the same direction, one should remember the preponderant role played by the measurement processes in the *operational definition* of concepts suggested by Bridgman (1927). From the psychological side, a great part of the so-called "cognitive psychology" goes in this direction.

[2] Throughout this study we would rather use terms such as "corporeal world", "body", "corporeal possibility", etc. instead of terms like "physical material world", "material thing", "matter", etc. because the former refer almost exclusively to the realm of the "composed" and "gross" manifestation, whereas the latter may refer to the notions of *materia prima* or *materia secunda* used by the Scholastics of the Middle Age after Aristotle. These *materiae* are to be distinguished from the conventional "matter", which is endowed with properties that enable it to be "sensible", while the former are no yet determined and remain as such "unintelligible" as a means of knowledge in the corporeal world. (This distinction may be regarded at first sight as pointless, but it will be very useful for future studies, in particular, when considering the *materia* of the corporeal world.)

Middle Age are important references as well.[3] As one goes along, one finds that most of these conceptions have, in radical opposition to the modern scientific method, a strong tendency to a "synthetic" and integral knowledge rather than to an "analytic" knowledge.[4] This perspective however does not necessarily avoid studying particular or concrete things. Indeed, all these doctrines take as a departure point an uncontradictable "whole", containing every possible point of view, which, as we will see, coincides with the Infinite in its etymological sense.

In this context, any particular study must be a development of either a given point of view or an "internally" coherent plurality of points of view selected from the infinite possibility; the more closely defined each standpoint is, the greater its degree of particularization. Otherwise, such study loses its deductive character and its logical validity, for logic needs an axiomatic basis. In this sense, any particular science constitutes a study of a particular aspect or aspects of the universe in its more general sense. Concretely, the Hindu doctrine for example has six main traditional points of view, one of which, called *Vaishêshika*, represents one of the aspects of the "cosmological" point of view. It is the branch of the doctrine that applies to the knowledge of things in distinctive and individual mode, considering them more directly for what they are in themselves[5] (while *Nyâya*, a second point of view, treats these things in terms of their relationship with the human understanding, *i.e.*, "logic" in an extended sense). To give a better idea, this point of view corresponds most exactly to what the Greeks, especially in the Pre-Socratic period, called "physical philosophy" or "natural philosophy."[6] This is precisely the point of view this study is intended to be based upon.

---

[3] Western "thinkers" like Leibniz, Schrödinger and others have in our opinion also very interesting points of view.

[4] At this point, it is worth mentioning that here we understand "analysis" in opposition to "synthesis." The analysis implies the separation of a whole into its component parts, and a process in development. All synthesis is instead something immediate, so to speak, something that is not preceded by any analysis and is entirely independent of it; it is "synthetic" knowledge which lets us grasp principles. In this sense, contrary to the current opinion, according to which analysis is as if it were a preparation for synthesis and leading to it, so much so that one must always begin with analysis, even when one does not intend to stop there, the truth is that one can never actually arrive at synthesis through analysis.

[5] The *Vaishêshika* point of view is, because of its distinctive character, an analytical point of view which has its counterpart in the *Sânkhya* standpoint (this makes another of the six different points of views in Hindu doctrine). The latter is also a "cosmological" point of view in the sense that the "Cosmos" is also its object of study. However, in this case the "Cosmos" is envisaged starting from the principles from which it proceeds and which determine it in its every mode. This is then a synthetical point of view since it grasps things in their principles. Interesting developments in this direction can be found in Guénon (1997, part III) and Kak (2003, 1-3). On the other hand, Ramsperger (1939) offers an interesting discussion on the grounds of the "positivist" and "realist" points of view, but we need to note that these considerations have a much more limited scope than the Hindu or ancient Greek perspectives can have (see following note).

[6] "Cosmology" and "natural philosophy" in this context should be understood not only as the study of "corporeal world" or "physical material world", but as comprising the entire "Manifested World" in its most extended sense, the "Manifestation" being the result of the "action" exercised by the *Natura naturans* on the *Natura naturata*, or, equivalently, the "action" of *Purusha* on *Prakriti* from Hindu doctrines. Taken in their extended sense, we could say that the "Manifested World" results from the "action" of the "essence" on "substance" or, in the Aristotelian terms, εἶδος (eidos) and ὕλη (hyle). Concretely, "cosmology" in this sense includes the study of the principle of any kind of existence (point of view corresponding to *Sânkhya*) and this existence in itself (point of view *Vaishêshika*), understanding here "existence" in its etymological sense (from the Latin *ex-stare*), *i.e.*, the being that is dependent on a principle other than itself, or, in other words, one that is not for itself its own sufficient cause. The mind, for example, can be seen, from a given point of view, as the first principle of the "ideal" world in its psychological sense; both are therefore comprised in the "Manifested World." This is to be distinguished from the much narrower conceptions like "by 'exist' we mean 'to have a spatial-temporal locus' (Miller 1946, 281)", often found when attempts are done to bring together modern science with the conceptions just mentioned. (We are, of course, not contrary to the "rapprochement" of these different conceptions. Quite the opposite, the present work is in part an attempt to do so. However, in spite of some tendencies or similarities, we believe that one has to be very careful when doing some kind of "parallelisms"; Yaker [1951], for example,

From the former considerations, it can be seen that before any further development the uncontradictable "whole" or infinite possibility has to be, as much as this is possible, expressed in one way or another. Our intention is to do so and then to derive some specific applications. Not less important to us, for the proper intelligibility of this work, it is to try to give, when possible, some examples. In particular, in the following sections we deal with the concepts of space and time (respectively, *dish* and *kâla* for the *Vaishêshika*) as conditions of the corporeal world (therefore considering them in a distinctive way, and not as a non-developed possible or possibility; see below). We believe that this approach is very advantageous in many respects in relation to those generally used in theoretical science, specifically in modern physics. It seems to us this is a way out to the inexhaustible process (because analytical) of knowledge which modern science develops almost exclusively.

## II.    The Total Possibility, Possibles and Compossibles

It is evident that the "Whole" in the universal and absolute sense, cannot be limited in any way, because it could only be so in virtue of something exterior to it, and if anything were exterior to it, it would not be *the* "Whole." Then, in this sense, the "Whole" can be identified with the Infinite. Indeed, the Infinite, according to the etymology of the term which designates it, is that which has no limits. Strictly speaking, it is that which has absolutely no limits whatsoever, excluding here everything that only escapes from certain particular limiting conditions while remaining subject to other limitations by virtue of its very nature. Take for example the space modality without boundaries. This kind of space, since it has non-spatial frontiers, escapes this limitation, but is still subject to the limitation resulting from its spatial nature. This means that even in the case of an indefinitely extended space,[7] it is not infinite, because it is still limited by the mere fact of being space and leaving outside all the non-spatial possibilities.

This takes us to another important distinction. When a *particular* whole (and not the "Whole" in its universal sense) is concerned, there are two cases to be distinguished from one another. On the one hand, a true whole is logically anterior to its parts and independent of them, whereas, on the other hand, a whole conceived of as logically posterior to its parts, of which it is merely the sum, and whose existence as a whole depends on the condition of actually being thought of as such. In some sense, it is we who give it this feature. The first case contains in itself a real principle of unity, superior to the multiplicity of its parts, whereas the second has no other unity than the one our thought attributes to it. An example of the former is precisely the space, which is anterior to its parts because extended, but still determined by its own nature; and an example of the latter is an arithmetical sum obtained by the addition of its parts, taken successively, one by one (see § III.2 for more details). It is important to observe moreover that the "Whole" must in any way be likened to a particular or determined whole; properly speaking, it is "without parts", because these parts would be relative and finite of necessity—even if capable of indefinite extension, as space—and so could have no common measure with it, for the "Whole" in this sense, being the Infinite, cannot admit any condition or determination; for every determination, of whatever sort, is necessarily a limitation by the very fact that it must leave something outside of itself, namely all other equally possible determinations. Besides, limitation presents the character of a veritable negation (*i.e.*, "privation" in the Aristotelian sense). To set a limit is to deny to that which is limited everything that this limit excludes, and consequently the negation of a limit is properly the negation of a negation, that is to say, logically, and even mathematically, an affirmation, so in reality the negation of all limits is equivalent to total and absolute

---

offers an interesting study on the attitude of the "Doctors" of the Middle Age and some rather contemporary scientists).

[7] An example is given by a space with hyperbolic geometry (see, *e.g.*, Misner *et al.* 1973).

affirmation, *i.e.*, Infinite. Expressed thus, it may seem that we are merely defining terms, but in reality we are just "setting" the notion of That which has no limits, that is, that of which nothing can be denied, and is therefore that which contains everything, that outside of which there is nothing. This idea of Infinite, which is thus the most affirmative of all because it comprehends or embraces all particular affirmations whatsoever—it implies them all equally—, can only be expressed in negative terms by reason of its absolute indetermination. This idea is a logical and ontological necessity, since if contradictions are present, how can we be certain anything is true?

In this sense, the "Whole" can be envisaged from the point of view of the (total) Possibility, which comprehends all particular possibilities. Indeed, a limitation of the total Possibility is, properly speaking, an impossibility: what is outside of the possible can be nothing but the impossible, otherwise we are not talking about the total Possibility. In addition, since the impossibility[8] is a pure and simple negation, a true nothingness, it can obviously not limit anything whatsoever, from which it immediately follows that the total Possibility is necessarily unlimited; it is thus an aspect, insofar as it is permissible to say so, of the Infinite.[9] On the other hand, particular possibilities are limited by the coexistence of other orders of possibilities, and thus limited by virtue of their own nature. From this relative and partial point of view, it is worth using the distinction made by Leibniz between "possibles" and "compossibles." Compossibles are nothing but possibilities that are mutually compatible, that is to say, whose union in a complex whole introduces no contradictions into the latter; consequently, the compossibility is always relative to the whole in question. Moreover, it is clear that such a whole may be that of the characteristics constituting all the attributes of a particular object, or that of an individual being, or again may be something far more general and extended such as the totality of all the possibilities subject to certain common conditions, forming thereby a certain definite order or determined whole. So, for example, Euclidean and non-Euclidean geometries are both possibles, but the union of them in the same space (or spatial condition) implies contradiction; but these two possibles are nonetheless also realizable. The different modalities of space to which they correspond can coexist in the integrality of the spatial possibility where each must be realized after its fashion. In more general terms, *manifested possibilities* imply the *effective development of these possibilities in a conditioned mode*, but these same possibilities in their non-manifested essence cannot be subject to such conditions. In this sense and coming back to the spatial possibility, every spatial possible has its proper existence as such, which is essentially inherent in it, and becomes unrealizable simply because other possibilities are currently being realized. In more general terms, one could say that a world is the entire domain formed by a certain ensemble of realized compossibles. Thus, these compossibles must be the totality of possibles that satisfy certain conditions characterizing and precisely defining that world. The other possibles, which are not determined by the same conditions and, consequently, cannot be part of the same world, are obviously no less realizable because of all that, but, of course, each according to the mode befitting its nature. Take as examples the "ideal" world in its psychological sense, that is, the world of thoughts and, on the other hand, the "corporeal" or the, usually called, "physical material" world. Every body or "material" object is subject to the spatial condition; thoughts, even though not being spatial,[10] are nevertheless realizable.

---

[8] This impossibility must not be confused with what is called the "physical" necessity or necessity of fact, which is the impossibility that beings and things could fail to conform to the laws of the domain to which they belong. This latter kind of impossibility has a relative validity and cannot be thus a "nothingness", the "physical" necessity being subordinate to the conditions by which the given "world" is defined, and which are valid only within the special domain concerned.

[9] These considerations are very closely related to the "non-dual" conception of the ultimate Reality found, in one way or another, in many doctrines throughout history: from the Eastern side we have for example Shankarâchârya (see, *e.g.*, Maharshi 1997, 174-78), Tchoang-tzeu (see, *e.g.*, Wieger 1950, 395-97 & 437-39) and Mohyiddin Ibn Arabi (1911); whereas from the Western side the "Presocratic tradition" is the main reference, from which the best to mention is perhaps the Pythagoreans (see, *e.g.*, D'Olivet 1991, 197-99 & Guthtie 1978, 246-51) and Parmenides (see, *e.g.*, Guthrie 1980, 26-43 & Coomaraswamy 2001, 63-64). Interesting considerations in this direction, but with a somehow more limited scope, can also be found in Kak (1999; 2004) and Narayan (2007).

It is doubtless always legitimate, if necessary, to envisage certain orders of particular possibles to the exclusion of others, and this is what any science actually does; but it is certainly not legitimate to affirm that this is the "Whole" or total Possibility, since, as such, sciences are in fact only restricted conceptions, which can have a certain validity in a relative domain by dint of some of their elements. Science is true so far as the world of science is concerned (as it is often said, every "system", "internally consistent", is true in what it affirms and false in what it denies). On the other hand, one could either consider one condition in isolation, as done above for the spatial condition, or, at a more restrictive point of view, but of the same order, consider the totality of the conditions that determine a world. It goes without saying that the several *conditions* thus united must be mutually compatible, and their compatibility obviously entails that of the *possibles* they include respectively, with the restriction that the possibles subject to the given group of conditions can only constitute a part of those which are comprehended in each of the conditions envisaged apart from the others. From which it follows that these conditions in their integrality, beyond what they hold in common, will include various prolongations.

In this sense, modern physical science studies some aspects of the "physical material world" that results from a set of *possibles* manifested under the union of a very particular set of compatible *conditions*. Here, we refer to this world as the "corporeal world", or "corporeal existence", and the Hindu doctrine designates it as the domain of "gross" manifestation (in opposition to the "subtle" or "psychic" manifestation). The corporeal world is indeed a possibility (otherwise how can we perceive it?) defined and delimited by some conditions; work is in progress in order to determine what is the minimum set of such conditions and what they are, but all the doctrines we mentioned before agree in saying that two of them are the *spatial* and the *temporal* conditions (it is also said, by the way, that both are continuous[11]). In the present article our intention is to study these conditions as realized in the corporeal existence.

## III.    Corporeal or "Physical" Space

### III.1   Some generalities

Following the reasoning given above, space can in general be conceived of as the condition where spatial possibilities can be realized or, in other terms, as what contains these possibilities. Space is the *container* of spatial possibilities. It is then quite evident that by definition only spatial possibilities can be realized in space; but it is no less evident that this does not prevent non-spatial possibilities from being equally realized outside of that particular condition that is space. If, however, space were infinite, as some claim, there would be no place in the Universe for any non-spatial-condition possibility since infinite implies having no limits whatsoever, and, logically, thought itself in the psychological sense, for example, would have to be excluded from existence, except as being conceived with a spatial extension. Some may disagree with this example. In such case, one can still consider as a non-trivial example the domain of "pure discrete quantity" (see note 10). Far from being infinite, space is only one of the possible modes of manifestation, time being another (see § IV). This is so, even in the largest conception of the spatial condition. Indeed, space may embrace the set of all particular spatial extensions, which can differ from one another by their number of dimensions or by other features. The ordinary "Euclidian" space, for example, is only a particular space, since it is a three dimensional extension fit out with a Euclidean geometry. This is not, however, the only conceivable modality of it. There are the "non-Euclidian" geometries, geometries of more than three dimensions and so on. The latter are a kind of the so-called geometric spaces, which contain geometric forms or figures.

---

[10] In our opinion the things which the psycho-physiologists determine quantitatively are not really in themselves mental phenomena, but only some of their corporeal concomitants. In this sense, electrical pulses from neurons can be seen as some of the corporeal concomitants of the mental activity, such as, thoughts which by themselves have no spatial character, though, like other phenomena, they develop in time. While on the subject of mental phenomena, it may be added that, once they are seen to be akin to what represents essence in the individual, we wonder to what extent it is useful to look for quantitative elements in them, and to what extent they can be reduced to quantity. By the way, is not pure discrete quantity another possible that it is not subject to the spatial condition? (The study of these and other concomitant phenomena regarded as a particular point of view of "natural science" in Islamic tradition constitutes another example of the conception of science mentioned above. "Physiognomic" is a very concrete example in this respect; see, *e.g.*, Youssef 1939, 27.)

[11] This attribute is intrinsically tied to the notion of space and time. However, to show it, one would have to go beyond the spatial and the temporal conditions (thus beyond the corporeal world), mainly by considering a "point" (thus extensionless) and its possible relations. These developments cannot enter into the scope of the present article. However, some indications are given in § III.2. (See also the considerations about *aṇu* in Narayan 2008 and specially Coomaraswamy 2001, Chapter I.)

Analogically, there is the *corporeal space*, by which we mean, on the one hand, as a possibility, the space capable of containing bodies and, on the other hand, once it is realized, the space actually containing bodies. The latter is our main interest in the present study. We prefer to use the term "corporeal" space rather than the usual terminology of "physical" space, that is, the space containing the "physical material world", because the latter is susceptible of a much larger application. The word "physics", indeed, may be taken to denote the natural science envisaged as a whole, as the study of the whole of Nature; that is to say, that even thoughts and their laws, for example, may have to be included in such science (*cf.* note 6).

In this work, corporeal space is regarded as a realized space, that is, the space that actually makes part of the corporeal world. In other words, space[12] is here taken simply as the "place" where the possibilities of corporeal order occur. Space is then the container of bodies *and* corporeal phenomena or, still another way of saying the same, space constitutes the "field" (*kshêtra* in the Hindu doctrine) within which corporeal manifestation develops. In this sense, bodies and the phenomena of which they are the support—the corporeal phenomena—constitute the content of space. There cannot be space without bodies (and vice versa), because this would imply a container without content, that is, something that cannot exist effectively. The relation of container to content necessarily presupposes, by its very nature as a correlation (*i.e.*, a reciprocal relation between two terms or notions), the simultaneous presence of both of its terms. Thus, the container and content are two correlated concepts (one cannot exist without the other, one implies the other). In particular, in this respect, the correlation between bodies and space should not be confused, as it often is, with an identity. On the other hand, this does not mean, however, that one cannot include them in a common point of view, as included in a "higher" level of possible (in the sense of § II). In other words, even if space and bodies are very distinct things, of different nature (at least insofar as we consider them with respect to the corporeal world), this does not mean that we cannot find a common principle from which we can explain them, as the fact that they are correlated gives some indication. To illustrate this, consider the following example: when one talks about a son, one is implicitly talking about the idea of a father and, vice versa, whenever one talks of a son, anybody will know that, unavoidably, there is (or was) a father. These two ideas (father and son) are inseparable. There cannot be a son without a father, nor a father without a son; but the two functions, the two beings, though belonging to the same species, are completely different (in some sense, one could say that they are complementary). Therefore, even if one function cannot exist without the other, they are completely different. What is more, in this case, they *have* to be quite different in order that there be a correlation between them.

---

[12] From now on and for simplicity, we use "space" as synonymous with "realized corporeal space", clearly distinguishing and characterizing others kinds of spaces when necessary.

This example highlights the correlation between two beings. Father and son coexist as beings (they are implied mutually and, in this particular case, they are complementary). On the other hand, it leaves out the idea of coextension[13] that is proper (but not exclusive, since *any* pair-container content presents it) to space and bodies. *Space is coextensive with bodies* (*cf.* Guthrie 1980, 33 & Narayan 2008, § 4.4). Indeed, according to the given definitions, strictly speaking, if one wants to talk about space, one has to necessarily associate it with bodies and their phenomena. The only function of space is to contain corporeal possibilities: it is the "field" available for the manifestation of these possibilities. Therefore, for example, if we say that between this article and the reader there is some space, then between this article and the reader there must have to be realized corporeal possibilities.[14] We arrive thus to the conclusion that space understood as the container of the corporeal possibility leads us to the idea of coextension of space with bodies. Evidently, if they are coextensive, they are coexistent with respect to the corporeal world. The concept of coextension is more restrictive than that of coexistence, and therefore, the former implies the latter. If something is coextensive with something else, they are necessarily coexistent. On the other hand, if they coexist, they are not necessarily coextensive (as in the example of father and son). These conclusions will be enriched by the considerations developed in the rest of the article.

The "coextensivity" of space with corporeal possibility is in relation with the continuous attribute of bodies. Indeed, this attribute can be understood as a participation in the extensional property of space: bodies are continuous because they are extended. Similarly, the continuity of movement, as well as the various phenomena more or less directly connected to it, derives essentially from its spatial character (see note 11 & § IV.3). Thus, the continuity of extension is ultimately the true foundation of all other continuity that is observed in corporeal nature. This does not mean, however, that there could not be discontinuity in the (corporeal) natural phenomena, in other words, "nature could make leaps" (*cf.* Wieger 1950, 139). Extension is then a very important property in order to understand the corporeal world and corporeal space. It is therefore worth saying some words about it and about some properties related to it.

## III.2  Extension, indefinitude, divisibility, smallest parts and final elements

*Extension* is, because of its continuous character, susceptible of increasing and decreasing indefinitely. Indeed, the idea of *indefinite* implies that of a development of possibilities, the limits of which we cannot actually reach, that is, a development that in itself and in its whole course always consists of something unfinished. This characteristic of indefinite leads some to call it infinite. However, the indefinite always implies a certain determination and limitation, this limitation and determination coming from the nature of the "object" under study itself. In this sense, the indefinite proceeds from the development of finite, a development extended until the limits are found to be out of our reach, at least insofar as we seek to reach them in a certain manner that we can call "analytical" (this does not mean, of course, that these limits might not be reached by other means; see below). As an example, take the sequence of natural numbers: here, it is obviously never possible to "stop" at a determined point, since after every number there is always another that can be obtained by adding a unit. This sequence is formed by successive additions of the arithmetical unit itself, indefinitely repeated, which is basically only the indefinite extension of the process of formation for any arithmetical sum. Here one can see quite clearly how the indefinite is formed starting from the finite (the sequence of natural numbers constitutes a multitude, formed from the unit, that surpasses all numbers), and how the indefinite is a development that in itself and in its whole course always consists of something unfinished, something analytically inexhaustible,[15] but not infinite. In this sense, it is easy to see that extension (thus space) is susceptible of an indefinite development. Take for example a line, indefinite in a single dimension. It can be considered to constitute a simple indefinitude of the first order. A plane surface, indefinite in two dimensions, and embracing an indefinite number of indefinite lines, is then an indefinitude of second order, and so on.[16] It can then be said that each spatial dimension introduces a new degree of indeterminacy to space, that is, to the spatial continuum insofar as it is subject to indefinite increase of extension and thus yields what could be called successive powers of the indefinite. It can also be said that an indefinitude of a certain order or power contains an indefinite multitude of indefinitudes of a lower order or lesser power. Therefore, as long as it is only a question of the indefinite in all this, these considerations, as well as others of the same sort, remain perfectly acceptable, because there is no logical incompatibility between multiple and distinct indefinitudes, which, despite their indefinitude, are nonetheless of an essentially finite nature, and which, like any other particular and determined possibility or possible, are, therefore, perfectly capable of coexisting within total Possibility, which is *alone* infinite, since it is identical to the universal Whole. On the contrary, the coexistence of several infinities, even if supposed to be of different degrees, suffices to prove that none of them can be truly infinite (*cf.* § II). In this sense, as it should be clear by now, the spatial possibility, even in all its generality, is still only one given possibility, indefinite no doubt, even indefinite to a multiple power, but nonetheless finite in the sense already explained. We see thus how the idea of indefinite helps to solve many apparent paradoxes. It also helps to understand the divisibility property of extension and to clear up the question of smallest parts and final elements of space, as we are just about to see.

---

[13] One could "translate" the coextension (in this case of temporal character) in the son-father correlation by saying that, for example, both (living) beings have to coexist in this world, something that, of course, is not necessarily the case.

[14] Evidently, we do not only refer to the well known empirical verification that there is air filling up this space. Fields, radiation, etc. (in the sense of modern physics), are included in the corporeal possibilities.

We observe that *divisibility* is a quality inherent to the continuous nature of extension. On the other hand, as long as there is extension, it is always divisible and one can thus consider its divisibility to be truly indefinite, its indefinitude being conditioned, moreover, by that of extension (its limitations can only come from this nature itself). Consequently, extension as such cannot be composed of indivisible elements, for these elements would have to be extensionless to be truly "indivisibles" and a sum (a union) of elements with no extension can no more constitute an extension than a sum of zeros can constitute a number, ("different from zero"). This is why ("geometric") points, being indivisible, are by that very fact without extension (that is, spatially null) and therefore, they can in no way be considered as constituting an element or part of length (or of a line). In fact, the true linear elements are always distances between points, of which the latter are only their extremities: points are limits or extremities, not constituent parts of linear elements.[17] From here, it is obvious that points, multiplied by any quantity at all, can never produce length; the true elements of a magnitude must always be of the same nature as the magnitude, although incomparably less. This leaves no room for "*indivisibles*" (or *smallest parts*) and it allows us to observe that ordinary quantities and infinitesimal quantities (in the sense of infinitesimal calculus) of various orders, although incomparable among themselves, are, nonetheless, magnitudes from the *same species*, that is, of the same nature of the "continuous" from which they arise. Thus, one cannot arrive at indivisible elements without departing from the special condition that is extension and the latter could not be resolved into such elements without ceasing to be an extension. To illustrate this, consider the following example: as the length of a circumference increases the further the circumference is from its center and one might at first suppose that it "contains" more points. Yet, if we reflect that every point in a circumference is the end of one of its radii, it follows that there are no more points in the greater circumference than in the lesser. Moreover, and this is what we wanted to emphasize, if there are always as many points (if it is possible to employ such a mode of speech under these conditions) in a circumference that diminishes as it approaches its center, then as this circumference is in the limiting case reduced to the center itself, the center, though being a single point, must contain all the points in the circumference, which amounts to saying that all things are contained in unity (in this case the geometric unity: the point). This example has the advantage of illustrating at the same time that, in spite of the fact that points are not constituent elements of extension, the point can be regarded, in some sense, as the principle of extension. The point is the "synthetic" principle of extension (*cf.* note 11).

---

[15] This is so from the point of view of analytic knowledge, but not from that of synthetic knowledge. It is in this sense that the limits of the indefinite can never be reached through any analytical procedure, or, in other words, that the indefinite, while not absolutely and in every way inexhaustible, is at least analytically inexhaustible (see below and § VI for more details). Analytically, "*the* atom is as inaccessible as *the* universe" (Malisoff 1939, 264).

[16] Here it is important to point out that we say "the surface *embraces* an indefinite number of lines", not that "it is *constituted* by an indefinite number of lines", just as a line is not composed of points, but rather embraces an indefinite multitude of them. In addition, this language is much more rigorous than common language even for closed, hence obviously and visibly finite surfaces and curves, which nevertheless contain, as it is usually said, "an infinite number of lines or points", respectively. One more precision is worth mentioning: if from a certain point of view one can legitimately consider a line to be generated by a point or a surface by a line, this essentially presupposes that the point, or the line, be displaced through a continuous motion, embracing an indefinitude of successive positions; and this is altogether different from considering each of these positions in isolation, that is, considering the points or lines as fixed and determined, and as constituting the parts or elements of the line or the surface, respectively (thus considering them analytically and hence giving rise, for example, to paradoxes as those from Zeno). Conversely, when one considers a line to be the intersection of two surfaces, and a point the intersection of two lines, these intersections must not, of course, by any means be conceived as parts common to the surfaces or lines; they are only limits or extremities of them (see below).

Therefore, the "continuous", thus, extension, insofar as it exists as such, is always divisible, and, consequently, it could not have smallest parts. "Indivisibles" cannot even be said to be parts of that with respect to which they are indivisibles. If we would like, however, to talk about a "minimum", one should understand it as a limit or extremity, not as an element. Not only is a line less than any surface, it is not even a part of a surface, but merely a minimum or an extremity (*cf.* note 15). One should now be convinced of the impossibility of getting within extension itself an indivisible or smallest part of extension (or space).

On the other hand, one could still ask about the "reality" of an "actual division" of extension or, in other words, the simultaneous existence of all the elements (or parts) of a given division of extension.[18] This idea amounts to supposing an entirely realized indefinite and on that account is contrary to the very nature of indefinitude, which, as we have seen in some detail, is always a possibility in the process of development, thus essentially implying something unfinished, not yet completely realized. Moreover, there is, in fact, no reason to make such a supposition, for when presented with a continuous set we are given the whole, not the parts into which it can be divided; and it is only we who conceive that it is possible to divide this whole into parts capable of being rendered smaller and smaller so as to become less than any given magnitude, provided the division be carried far enough. In fact, it is consequently we who realize the parts, to the extent that we effectuate the division. Thus, what exempts us from having to suppose an "actual division" of extension is the following: a continuous set is not the result of the parts into which it is divisible, but is, on the contrary, independent of them (it is a particular whole superior to the multiplicity of its parts, see § II), and, consequently, the fact that it is given to us as a whole by no means implies the actual existence of those parts. All continuity, when taken with respect to its elements, embraces a certain indefinitude. In this sense, these elements have a "conventional" character, *i.e.*, it is we who give a reality to the elements or parts as such, by an ideal or effective division. In this respect, consider the following example: the "number" 2 regarded, not as an arithmetic quantity, but as the geometric magnitude consisting of two units distance from a point arbitrarily chosen to play the role of origin. This magnitude can be conceived of, as it is well known, as the sum of a geometric series, $1+1/2+1/4+1/8+\cdots$; however, each addend or summand is of the same nature as 2, that is, all of them are continuous magnitudes (segments marked off by lengths 1, 1/2, 1/4, etc.), as little as they may be, and even if they decrease indefinitely. It is worth mentioning that 2, taken in this way ("geometrically"), is a continuous set constituting a "true whole" in the sense that it is logically anterior to its parts and independent of them. When representing the "number" 2 as the "sum" of an indefinite multitude of "addends" (or intervals), we have *arbitrarily* divided it into parts, which does not imply in any way the previous and effective existence of these parts; here, it is we who give a reality to the parts as such. On the other hand, and in an inverse sense, when taking the number 2 arithmetically (as a discontinuous quantity), it does not constitute a true whole, since it is the simple sum of its parts (two times the number 1, the arithmetic unit), and, consequently, is logically posterior to them. As such, it is nothing other than a "being of reason" or *ens rationis*. It is "one" and "whole" in so far as we conceive it as such. In other words, by itself, the number 2, conceived arithmetically, is strictly speaking only a "collection", and it is we who, by the manner in which we envisage it, confer upon it in a certain relative sense the character of unity and totality.

---

[17] A similar reasoning could be applied to curves with respect to surfaces, surfaces with respect to three dimensional space, and so on.

[18] In some branches of modern sciences these parts are referred to as "final parts", on the supposition that they are an "actual", "effective" or "fundamental" division of extension.

### III.3  Some conceptual considerations about space in modern science

It can be objected that the previous reasoning can be applied to geometric space but not to corporeal space. In some sense, this is indeed the case: the characteristics of the geometric space cannot purely and simply be transferred to the realized corporeal space. As we have said in § II, one could either consider one condition in isolation, as done above for the geometric spatial condition, or, with a more restrictive point of view, but of the same order, consider the totality of the set of mutually compatible conditions that determine a world. Their compatibility obviously entails that of the possibles they include respectively, with the restriction that the possibles subject to the given group of conditions can only constitute a part of those taht are compromised in each of the conditions envisaged apart from the others. From which it follows that these conditions in their integrality, beyond what they hold in common, include various prolongations. In this sense, while it is perfectly conceivable to have, for example, a closed curve within the geometric possibilities, this cannot be the case in the corporeal possibility. In the corporeal existence, because of the effect of the combination of spatial and temporal conditions, everything is in movement; as a consequence, it is materially impossible to draw effectively a curve in space which is truly closed. For example, if we want to draw a circumference beginning in a given point of space, because of our movement, at the end of the drawing we will necessarily end up at a different point from that of departure. We never come back again to the departure point.

We can thus say that in the corporeal world, notably because of the presence of the temporal condition, geometric figures appear only approximately. Therefore, when studying, for example, the geometric space in isolation, the obtained results can be applied to more qualitative spaces, like the realized corporeal space, but restricted to the additional conditions. The latter inherits the properties of the geometric space compatible with the set of conditions of the corporeal possibility. On the contrary, if geometric space does not present some properties, this does not necessarily imply that these features are also absent in the realized corporeal space. In what follows, we show how this kind of considerations may help to clarify many points in modern science.

One of them comes from an objection that has been addressed to us with respect to the assertion that "points can in no way be considered as constituting an element or part of length." This objection can be formulated as follows: there exists a branch of modern mathematics called "measure theory", which claims that a non-zero length, of a given interval of a line, can be obtained from a *countable sum of points*, provided that an *appropriate* notion of "measure" is defined.[19] However, it is not without reason that this theory is sometimes called "abstract measure theory" (see, *e.g.*, Reed 1992, § I.4). Indeed, in such a theory, one is able to define the "measure" of a point to be different from zero (a trivial example is given by the "Dirac measure", which, in particular, assigns the number 1 to a given point $x$ and 0 otherwise)! Thereafter, it is not surprising that one is able to obtain a non-zero length from a countable sum of points. Nevertheless, this "measure" has nothing to do with the realized space in the corporeal world—or even geometric space, in its most common conception, the spaces herein under study. Therefore, this objection is ill-defined within the considerations discussed in this work. Otherwise, if somebody would insist on establishing a relationship between this "measure" and the corporeal world, it would amount to asking (in "naïve" terms): what is the ruler with which we would obtain a non-zero outcome when "measuring" a point?! (That is, a "point with size.") This is actually a misleading question, because in reality a point cannot be measured: it escapes the spatial condition. In addition, the "physical" measurement, *i.e.*, the notion of measurement within the corporeal world, is principally concerned with the domain of continuous quantity, that is to say, it is concerned most directly with things that have a spatial character (*cf.* JCGM 2008). Indeed, this amounts to saying that measure is specifically concerned either with space itself or with bodies (by reason of the spatial character they have). In the first case, the measure is more "geometrical"; in the second case, it would more usually be called "physical" in the ordinary sense of the word. But, in reality, the second case leads back to the first, since it is only by virtue of the fact that bodies are situated in space and occupy a certain defined part of it that they are directly measurable (but, recall, bodies cannot be only "spatial"), whereas their other properties are not susceptible of measurement, except to the degree that they can in some way be related to space (while some reflection on the subject may suffice to become convinced of the latter claim, it is our intention to develop it in future studies, and in the following section we present some indications related to the measurement of time). Finally, it should be stressed that the definitions of "abstract measure theory" are based on "abstract" generalizations of this "intuitive" notion of measure and size. Otherwise, why should one talk about "defining an extension [generalization] of size" for which, by the way, it is we who decide "which sets are to have a size" (see, *e.g.*, Reed 1992, 14)?!

On the other hand, observe that, essentially because of the "coextensity" of space with corporeal possibility, there cannot exist indivisible corporeal elements, without falling into contradiction. Indeed, these kinds of elements would have to be without extension, and then they would no longer be corporeal, since, by the very definition, the word corporeal necessarily entails extension,[20] although this is not the whole nature of bodies (see below).

---

[19] Grünbaum (1952) gives a concise study on the subject. Without being experts on the matter, it seems to us that such study offers a mathematical ground to many of the "verbal" considerations herein developed. There should be, therefore, complementarity, and not contradiction, among both perspectives.

[20] There are no indivisible corporeal "particles", or "atoms" in the sense given by some ancient Greek philosophers, or by some Hindu and Buddhist schools, which, by the way, were regarded as more or less "heterodox" in this respect (see, *e.g.*, Shankarâchârya's comment on the Brahma-sûtra II.1.29 in Bâdarâyana (1904, 354) for more considerations on the "Indian atomism," see Coomaraswamy (2001, chapter I) and references therein). In the same sense, there are not indivisible fractions that cannot yield ever smaller fractions in the numerical order, or linear elements that cannot be divided into ever smaller elements in the geometric order. Furthermore, as Coomaraswamy (2001, 18) claims, "it cannot be too strongly insisted that in

Now, one can still consider, as in the case of geometric space, an "actual division" of corporeal world or of bodies; but these parts, no matter how small they are, cannot have a "fundamental" character: they are at most a theoretical agreement or an empirical verification. In the case of the empirical verification the strongest claim we can make is that "in the present state no division has been observed of a given body." It is in this sense that some theories in modern physics give a "fundamental" character to the so called "ultimate" (or "smallest") parts of "matter", better known as "elementary particles"[21] (see, *e.g.*, Redhead 1980) and, to a smaller extent, some other theories focus on the "ultimate" (or "smallest") parts of space.[22] However, in this context, "ultimate" or "smallest" should be understood in an "effective" way, meaning that there is no current way for modern physics to go beyond this "fundamental structure of matter"[23]. It is then we who, for any given reason, give some importance to these "parts." They *may* be useful, to a given extent, to describe or "represent" at some level the corporeal world; they are "pieces" (portions) of the same nature as the whole, which we decide to study analytically, namely, by dividing the whole into parts.[24] In this respect, we believe that there are no solid arguments leading to the claim that these kinds of theories are penetrating to the essence (the nature) of bodies and/or space (*cf.* §§ IV.2 & VI). One could say instead that in this context one *may* be getting into a more detailed "analysis" of the whole under study, which does not imply, however, that we are getting into something essential (in its "Platonic-Aristotelian" sense (see, *e.g.*, Kaminsky 1957). The parts of a "continuous" (or any "true whole") are of the same nature (of the same species) as the "continuous" itself; the "continuous" or any chosen part of it, because of its very nature of divisibility, is analytically inexhaustible. Coomaraswamy (2001, 62-63), in order to distinguish between the synthetic and the analytic points of view (*cf.* notes 4 & 15) in this context, put it this way: "Our concern is only with the really and absolutely indivisible and undimensioned atom or point that gives a meaning to time or space ('non-spatial and non-temporal intuition is the condition of the interpretation of the space-time world itself'), and not at all with such 'atoms' as have now been actually 'split', or with those of the 'atomists' such as Leukippos for whom 'there are an infinite number of them, and they are invisible owing to the smallness of their bulk'; atoms

---

the traditional doctrine everywhere time and space are uninterrupted continuities, and that were it otherwise the hare would never succeed in passing the tortoise. It is only when time is thought of as discontinuous that the operation of causality becomes unthinkable."

[21] The term "fundamental building blocks" is also of common use, meaning: "objects that are simple and structureless—not made of anything smaller." This concept differs, of course, from that one used in physics in a non-"quantic" sense: for example, in Newtonian and relativistic mechanics. In the latter, a "particle" is simply a body whose dimensions are so small, compared to the distance involved in its movement, that they can be ignored when studying its movement; this is, by the way, an example of the "idealizations" discussed in § IV.2.

[22] Most of the theories going in this direction are the so called "quantum theories of space-time" or some interpretations of "quantum gravity" (see, *e.g.*, Rovelli 2004; 2006). By the way, according to some of the interpretations of these theories, there is a "fundamental" *determinate* space length; for example, Planck length, $l_p$=1.6x10$^{-35}$$m$, is considered by Rovelli (2007; 1289) as a possible "minimal length", as a measure of the "granularity" of space. Furthermore, since these theories treat time and space jointly, the discretization of time is also therefrom entailed (see §§ IV.3 & V).

[23] A rather informal but "official" (because conceived by the Particle Data Group) illustration of this claim can be found in the "educational" presentation of "The Standard Model" given in the site: http://www.particleadventure/index.html. On the other hand, these "effective" limits are sometimes seen as a consequence of the "indeterminacy principle" in quantum mechanics; see, *e.g.*, Fitch (1936).

[24] Malisoff (1939, 262) put it in very clear terms: "All scientific systems are subject to an atomistic analysis, meaning thereby that when properly analyzed they can all show something which will play the rôle of an indivisible (relative to that analysis). The 'atomistic' analysis will lead either to electrons or quanta or simple ideas or sensations or... entities of any magnitude, provided they are themselves taken [!] as structureless. The above proposal could be made to cover cases where the atom (considered as a structure in a *different* context) could still be genuine atom in the sense of being the indivisible of an atomistic analysis. Thus old, valuable portions of science can be conserved in their sphere of usefulness without being 'wrong'. An instance of this is the kinetic gas theory for an ideal monatomic gas. The atoms of such a gas are indivisible within the scope of or for the purpose of this type of atomistic analysis."

that are 'not mathematically indivisible', but each of which '*has magnitude*' and extension, and of which, therefore, perceptible things *can* be constituted,—atoms that can, in fact, only be so called for so long as men have not yet been able to divide them, and which are really only *particles*: and can therefore, quite logically, be thought of as constituent parts of great magnitudes. Atomic constitution implies, indeed, a discontinuity of matter, but does not require a discontinuity of the space in which they must be thought of as arranged, nor does it require that this space should be literally a void. All traditions speak of an original separation of heaven and earth, in order that there may be a room or space in which things can exist; but the space thus created is aerial rather than empty."

So far we have assumed that bodies take up space, which means that bodies are conditioned to space. Otherwise, how could they be extended or have a volume? This does not mean, however, that the all nature of bodies can be reduced to that of space. Indeed, as we have already argued, space and bodies are two distinct things. It is true that bodies are extended because they participate in the extension of space. Who says corporeal says necessarily extended. But, contrary to what Descartes and other partisans of mechanistic physics have pretended, the extension does not constitute at all the whole nature or essence of bodies, that is to say, the whole nature of bodies cannot be reduced to extension. To say that a body is nothing but extension in a purely quantitative sense[25] is really the same as to say that its surface and its volume, which measure the portion of extension actually occupied by it, are the body itself with all its properties. Some of these properties (*e.g.*, body's size) are certainly taken from the spatial condition, as well as other determinations, such as, the situation of bodies in space. However, some other conditions are required to define a body, including time (see § IV). In this respect, but at the other extreme, some interpretations given to the theory of general relativity establish that space is in all respects determined by bodies (see § V). Now, in this work we have never stated that space is a (logical) consequence of bodies, but, as previously indicated, it is a matter of correlation (container-content, in this case) which gives coextension as a result; and, in any case, even for the theory of relativity it is not the existence of space which is determined by bodies, but its "shape" (its geometric structure, see § V).

Between these extreme positions: *a)* bodies are entirely explainable from the space nature and, the contrary, *b)* space's nature and its features are reducible to those of bodies, there are many intermediary standpoints. In particular, there is rather a tendency to deem the existence of space as *independent* of that of bodies, as the following usual kind of statement shows: Let's study the space without bodies; then, once we have the space under 'control', let's study bodies; once the bodies under 'control', as a final step, let's put bodies in space and see how everything works together (*cf.* the "Newtonian mechanics" of undergraduate books, *e.g.*, Feynman 1963, 2-3). From here, one can easily consider portions of "empty space" between bodies or even a completely "empty space", as if the space were already given before introducing bodies into it. It is clear to us that some approximate conceptions—such as an empty space with bodies inside in this case—may have some practical and even theoretical utility, but this is not here our concern. Instead, we now explain why an empty space cannot be a possibility of the corporeal world.

The idea of an empty space is very closely related to that of a homogeneous space, the latter also being a very common "idealization" in modern science, by the way. In particular, it is a basic input—or supposed so—in many theories of modern physics (these include all theories using Euclidean space, such as Newtonian mechanics and special relativity).

---

[25] There have been more sophisticated attempts to treat space (or space-time, see § V) as the primary concept from which the properties of bodies can be deduced (*e.g.*, Misner et al. 1957). These approaches, in general, take into account more qualitative aspects of space.

**Homogeneous space.** A homogeneous space can only be conceived of as a space without bodies, and even without geometric figures,[26] or, otherwise, as a space with the same (and identical) body everywhere (for example, in modern terms, "a homogeneous field"). Clearly, none of these possibilities occur in realized corporeal space, since it contains a diversity of bodies. The idea of a diversity of bodies can be sustained by, besides the evidence of this fact, the principle of indiscernibles discussed in § IV.2, and the considerations therein developed. In this sense, the mere presence of bodies suffices to determine qualitative differences between the parts of space they occupy. In addition, if we consider the size (the magnitude) as the quantitative aspect of space, and, on the other hand, we observe that a homogeneous space presents the unique property that its parts are indistinguishable one from another by any characteristic other than their respective size, then, a homogeneous space would amount to a space denuded of any qualitative element, that is, a "purely" quantitative space;[27] but, again, because of the presence of bodies, this is clearly not the case of realized space in the corporeal world. Therefore, the homogeneity cannot be a property of realized "corporeal space", as it is considered by the physical theories just mentioned. On the other hand, homogeneous space can be conceived of as *the limit* of a process where bodies are, so to speak, "absorbed" in a substance containing potentially all the corporeal possibilities; but even this "primordial" substance[28] needs a container, a "space", to be manifested, and vice versa, as we are about to argue.

**Empty space.** An empty space, that is, a space without any kind of body (or even any corporeal possibility) is of a very common use,[29] but strictly speaking, this cannot be seen as more than an approximate conception, because space is coextensive with the corporeal possibility. In other words, the conception of an empty space would lead back to the conception of a container without content, but this is something that we have discarded from the very beginning: the relation of container to content necessarily presupposes, by its very nature as a correlation, the simultaneous presence of both of its terms. Moreover, an empty space can be conceived of as a space denuded of any qualitative aspect ("qualitative emptiness"); but, clearly, this can have no place in the corporeal world, since it presents a multitude of qualitative aspects. From the previous paragraph and these considerations, we conclude that the hypothesis of an empty space implies the hypothesis of a homogeneous space (except for the limit case previously mentioned).

**Space with holes or holed space.** This space is usually conceived of as a space with "holes", where there is not any kind of body ("empty holes"). What we have said about the coextensivity of bodies and space should be enough to dispel this idea: when we talk about space, we have to necessarily relate it to bodies. Otherwise, we insist, we would be forced to bring out the hypothesis of portions of space devoid of any qualitative element (*i.e.*, portions of space fitted up solely with the size aspect), which are immediately rejected by the considerations offered above. In fact, each of these portions is a particular case of the empty space considered above.

---

[26] In general, figures would introduce a heterogeneity to space (recall the space containing only pure geometric forms is the geometric space).

[27] To this the objection might perhaps be raised that there exists the hypothesis of having a "homogeneous field" spread all around the space (as some cosmological modern theories do). But if this is a homogeneous field in all respects (in order to keep up the homogeneity hypothesis), what are then the qualitative aspects brought about by it? (see below).

[28] See § V.1 for more considerations about this "substance." In any case, notice that we have talked of a limit, so that homogeneous space is not to be confused with the realized (corporeal) space itself. Otherwise it is like confusing the circumference whose radius diminishes indefinitely with its center, a point. This point is the limit of the circumferences in this way generated. It is therefore outside such succession (*cf.* § III.2).

[29] Even in the theory of general relativity, the concept of empty space seems to be accepted: the Schwarzschild solution, for example, is a "vacuum solution with a spherical symmetric source". Does this mean that the space is filled up with "gravitational field"? Would it then still be appropriate to talk about "empty space"? (for more considerations about this we refer the reader to § V).

Therefore, *realized corporeal space does not extend beyond the corporeal world, and inside the latter, bodies (better to say, corporeal possibilities) are spread out everywhere*. These considerations, as well as the following, have an epistemological character in the sense that they aim to better understand the methodological approaches of some branches of modern physics and their limits. In this sense, before closing this section and trying to foresee some possible objections, we would like to say that the so called "quantum effects", as far as considered as measurable corporeal phenomena, do not change the previous considerations in the sense that these effects are necessarily part of the "corporeal possibility" and thus contained in the corporeal world. (Some work is already in progress in this direction.)

## IV.    Time or Corporeal Duration

### IV.1  Some generalities.

We consider time as the container of corporeal events, and everything that has been said about the correlation of container and content with respect to space and bodies and the consequences of this correlation could be repeated with respect to time and events. However some precisions deserve to be made for the better intelligibility of the concepts "event" and "time."

First of all, an event in this context should not be confused with that one used, for example, in a more restrictive and technical way, in some areas of modern physics (notably in the theory of relativity), where an event is defined as a given point of "space-time", without spatial extension nor time duration (see § V). The notion of "event" with respect to time is analogous to that of body with respect to space; just as this one extends in space, an event is a happening extended in time (it has a duration: an event always *lasts* some time). With regard to time, an event must be understood in its notion of development, that is, how the event unrolls (and not in its more general meaning of a "possible" not yet manifested, *cf.* § II). Therefore, the temporal equivalent of the geometrical point is an *instant*, that is, something that has no duration at all. Instants conceived of as indivisibles are no more parts of duration than are points of space. Instants can in no way be considered as constituting an element or part of duration. Instants are limits or extremities of a duration. In this respect, one has to always keep in mind that an event has a (temporal) duration[30] and that there can no longer be any succession within an instant. All this is not in contradiction, as in the case of space, with the idea that the instant can be seen as the "synthetic principle" of time (*cf.* § III.2 and, *e.g.*, Coommaraswamy [2001, chapter I]).

On the other hand, the nature of time is much more difficult to grasp. Space can be measured directly; time, instead, can only be measured by relating it back to movement, thus, we can say, to space (*cf.* Narayan [2007, § 5.1]; Biser 1952). Evidently, as for any continuous magnitude, we measure a duration (or time interval) by using an arbitrarily chosen interval as time pattern. However, the "unit" of time (or pattern) requires for its conception the notion of movement, and, therefore, the notion of space.[31] It should be clear that the direct inaccessibility of time through measurements does not imply, in any sense, that time does not exist or that we may reduce its nature to that of space.[32] In this respect, there are some approaches in modern physics that consist of considering evolution of bodies with respect to themselves, that is, the description of the changes (in general) of some given bodies with respect to the changes of some other bodies. This is, to some extent, an interesting idea when trying to *describe* (or "*represent*") evolution of bodies in the corporeal world. However, in our opinion, this does not permit saying that we are getting around time, or that time does not exist at all, or again that we are penetrating the nature of time.[33] This would be like a geometer believing that he is exploring the nature of (geometrical) space when doing geometry; the reality is that space and figures are presupposed by him. In the same sense, we wonder if, by exploring such theories, which consider *exclusively* the phenomena happening in the corporeal world, we can access the nature of space and time. The corporeal world is already a result of the combination of space and time (plus other conditions). In this sense, we would like to insist on the fact that we do not deny the possible usefulness of these procedures; but it seems unjustifiable to abide with some of the interpretations of such theories, such as, that movement (or change in general) has its principle in the movement itself. Just as the extension is contained "synthetically" in the spatially null nature of the point (see § III.2), movement, as Aristotle affirms, must be contained "synthetically" in the "motionless motor" (*cf.* Metaphysics, Book XII, parts 7-8). The prime mover of all things must be itself motionless (this is, by the way, one, among others, of the meanings of the Far-Eastern "*wu-wei*"; see, *e.g.*, Wieger 1950, 255, 397). From the ontological point of view Mueller (1943, 153) put it in these terms: "Everything changes, evolves, becomes. But every changing, evolving, becoming process also IS. If everything changes, then this is eternally true. The truth of change is its changeless being [!]. True being is absolutely identical with itself and is in no manner disrupted by becoming." (*Cf. Timaeus* 28-29 and 52 by Plato.) Therefore, any action or change is in agreement with the logical and ontological conditions given by this changeless being.

---

[30] In ancient philosophy, in particular, the Scholastic philosophers, the term "duration" was used to characterize all possible modes of succession, that is, in short, any condition that could correspond analogically to what time is in the corporeal world. In this sense, time is a particularization of "duration"; time is only a mode, among many others, of duration. In scholastic terms, in this work, we are dealing with *tempus* and not for example with *aevum*, a more general mode of duration susceptible of a greater degree of indefinitude (*cf.*, *e.g.*, Aquinas [1988, first part Q. X]; Porro [2001, 131-32]). It is in this sense that sometimes we refer to time as the "corporeal duration." In that respect, it is worth saying that the Hindu doctrine developed also a very complete and complex science related with (cyclic) "durations" of greater or lesser degree of indefinitude; there are for example, in "decreasing" order: *kalpas, manvantaras* and *yugas*; the latter being very closely related to what we call here "corporeal duration" (*cf.*, *e.g.*, Tilak [1903, chapter XIII]).

However that may be, it is evident that periods of time are qualitatively differentiated by the events unfolded within them, just as the parts of space are differentiated by the bodies they contain. It is not, therefore, in any way justifiable to regard, as being really equivalent, durations of time that are quantitatively equal when they are filled by totally different sequences of events. However, the symmetry between space and time in this respect is also not perfect, because the situation of a body in space can vary through the occurrence of movement: whereas that of an event in time is rigidly determinate and strictly "unique" (in the sequence of events), so that the essential nature of events seems to be much more rigidly tied to time than that of bodies is to space. This implies that in a sequence of events there is a succession, an ordered development or evolution. We call *temporal succession* the succession of events ordered on time: in other words, to the order followed by the content of time[34] (*cf.* Kak 2003, 15-16). There is still another remarkable difference between space and time, namely, the absence in the case of time of a quantitative science of an order comparable to that of the geometry of space (this should not be confused with the geometrical representation of time which is treated in what follows).

---

[31] We could go even further by claiming that we never measure a duration, but instead the space covered throughout this duration in the course of a movement for which the law is known; and as any such law expresses a relation between time and space, it is possible, when the amount of the travelled space is known, to deduce therefrom the amount of time taken to travel the given distance. Whatever may be the artifices employed, there actually seems no other way than this whereby temporal magnitudes can be determined (see, *e.g.*, Schlegel 1948, 25-7; Moulyn 1952; Moon *et al.* 1956). To seriously consider this affirmation, we would require a more detailed study on the nature of movement, which is not undertaken in the present work. However, we can say that it is in this sense that the earth's daily motion was used to define the "unit" of time. Presently, the "unit" of time consists of taking a certain number of periods of electromagnetic field's oscillations associated with the radiation emitted by a Cesium isotope in a given atomic transition (see Arias 2005).

[32] Time is not the only magnitude that should be related back to space in order to be measured. In fact, it seems that almost any property of bodies, in order to be measured, should be related back to space (more precisely, to its extension). This means that these properties can only be indirectly measured through space. Again, this *does not* mean that these properties are essentially spatial. This means instead that they can only be expressed, in one way or another (as time does through movement), in spatial terms.

[33] These approaches are generally comprised in the generic term "timeless physics theories" (see, *e.g.*, Barbour 2000; Rovelli 2004; 2006), but, in general, when one says "timeless" in this context, one means more particular and technical things. To give an example: one is said to have a "timeless" theory when one manages to get rid of a parameter which was, somewhat conventionally, called "time" (see also § V and Mondragon [2010, Part I, chapters 1 & 2]). Timeless in this sense does not mean, therefore, to go beyond the temporal condition, *i.e.*, for example, to get back to a single instant, the "now", the whole temporal succession (*cf.*, *e.g.*, Coommaraswamy [2001, 1-7 and 15-16]).

Now, among the above considerations, there is one subtlety to be inserted that minimizes the overall differences between bodies and events. When we say that a "given body" is displaced through space from position A to another position B, in reality there is one body in position A and another *different* body in position B. This can be easily understood by reflecting, for example, on the fact that what we conceive as a body in position A necessarily undergoes some modifications when being displaced to position B, therefore, strictly speaking, ending up as a different body. Somewhat conventionally we say that "the same body" is at position A and then at B, whereas in reality we are talking of different bodies,[35] perhaps, it is true, comparable one to another in certain aspects, nonetheless not identical (this is, of course, also true for a body in rest with respect to a given observer; see below for more details). These considerations may seem too trivial, but in the following we see how such "conventions" may have important consequences on the way human beings conceive the corporeal world, in particular with the advent of probability and statistics.

### IV.2  Repeatability of events and statistics

---

[34] We could define a succession of positions (on space) of a given body. However, as we have said, this succession could be modified by the effect of movement. It seems therefore not as "fundamental" as the temporal succession (see below).

[35] Schrödinger (1996, 131) put it in these terms: "If I observe a particle here and now, and observe a similar one a moment later at a place very near the former place, not only cannot I be sure whether it is 'the same', but this statement has no absolute meaning." In this respect, he found very enlightening the following words of Coomaraswamy (2001, 3-4): "The words 'real' and 'thing' have an interest of their own. 'Real' is connected with Lat. *res*, and probably *reor*, 'think', 'estimate'; and 'thing' with 'think', denken. This would imply that appearances are endowed with reality and a quasi-permanence to the extent that we *name* them; and this has an intimate bearing on the nature of language itself, of which the primary application is always to concrete things, so that we must resort to negative terms (*via negativa*) when we have to speak of an ultimate reality that is not any thing [as we have tried to do, by the way, in § II]. That a 'thing' is an appearance to which a name is given is precisely what is implied by the Sanskrit and Pali expression *nāma-rūpa* (name, or idea, and phenomenon, or body) of which the reference is to all dimensioned objects, all the accountable individualities susceptible of statistical investigation [see § IV.2]; that which is ultimately real being, properly speaking, 'nameless'. 'Name-and-appearance in combination with consciousness are to be found only where there are birth and age and death, or falling away and uprising, only where there is signification, interpretation, and cognition, only where there is motion involving a cognizibility as such or such' (*Digha Nikâya*.2.63). The Vedantic position is that all differentiation (naturation or qualification) is a matter of terminology (*vâcârambhaṇaṁ vikâraḥ, Chândogya Upaniṣad* 6.1.4-6, cf. *Saṁyutta Nikâya* 2.67 *viññânassa ârammaṇam*); and the same way for Plato, 'the same account must be given of the nature that assumes all bodies; one cannot say of the modifications that *are*, 'for they change even while we speak of them', but only that they are 'such and such', if even to say that much is permissible' (*Timaeus* 50 A, B). In this passage, the 'nature referred to is that primary and formless matter that can be informed,... nature as being that by which the Generator generates' (Damascene, *De fid. Orth.* 1.18) or 'by which the Father begets' (St. Thomas, *Sum. Theol.* 1.41.5)."

Nowadays, both in daily life and science in general, there is an increasing use of, and an ever growing importance attached to, the notion of repeatability of events (or phenomena), statistics, and probability. We think it is worth saying some words about this, since it is directly related to the notion of event and the qualitative aspects of time.

In modern science, a basic assumption and requirement is that one can repeat the same ("identical") experiments as many times as one wants. This assumption is sometimes called the "principle" of repeatability or reproducibility of identical events or phenomena and, mainly because of its practical success, it is very widely accepted in the scientific community without further inquiry.[36] We believe that this "principle" has more to do with the inductive aspect of the scientific method and generally accepted conventions than with a true principle. It has to do, in fact, with the empirical and statistical character of modern science. In any case, this "principle", by using the "principle of indiscernibles", is something in reality inconceivable. Indeed, if by this principle we understand that there cannot exist anywhere two identical beings, that is to say, two beings alike in every respect, it can be said that if two beings are assumed to be identical in this sense, they would not really be two, but, as coinciding in every respect, they would actually be but one and the same being (*cf.* Forrest 2010). In order that beings may not be identical or indiscernible there must always be some qualitative difference. This principle carries with it the absence of any repetition of particular possibilities, which in its turn is a consequence of the infinite "aspect" of the total Possibility introduced in § II.

Moreover, the impossibility of having identical events automatically prevents us from accepting the widespread assertion that *the same* causes always produce the same effects, since the happening of the "same causes", and that of "same effects", implies the repeatability of events. Indeed, one should say, instead, that *causes that are comparable one to another in certain aspects produce effects similarly comparable in the same aspects*. But, as is often forgotten, alongside the resemblances, which can, if desired, be held to represent a kind of partial identity, *there are always and inevitably differences*, because of the simple fact that there are, by hypothesis, two distinct things in question and not one single thing. In this respect, one has to be careful, because in some cases a superficial and incomplete observation might give the impression of a sort of identity; but actually differences can never be eliminated within the total manifestation.[37] Even if there were no differences left other than those arising from the ever-changing influence of time and place, they could still never by entirely negligible, because space and time cannot be, contrary to some conceptions in physics, merely homogeneous containers. On the contrary, temporal and spatial determinations—regarded as realized conditions of the corporeal world—have also in reality a qualitative aspect.

However that may be, it is legitimate to ask how, by neglecting differences, as if we refuse to see them, we can still possibly claim that our sciences are "exact." In reality, modern science (except for pure mathematics) is not much more than a series of approximations which is more or less suitable for some experimental or practical purposes, or particular facts, which are what this science is mostly interested in. This is not only true from the point of view of the application of science, in which everyone is forced to recognize the inevitable imperfection of the means of observation and measurement, but even from a purely theoretical point of view as well. Theoretical modern science resorts very often to over "idealized" models and suppositions that make one wonder about their real relationship with the situation under consideration.[38] In that respect, some representative examples in theoretical physics are a "massive material *point*" or an "*unstretchable* and *weightless* thread" or a "*free* falling body" or even a "*perfectly* isolated system." In more recent modern physics such suppositions have become more "sophisticated": there are "*identical* particles", "*auto*-interactions", and so on.

---

[36] This affirmation may be surprising for those who are aware of the existence of studies on the replication of experiments and its degree of validity. However, most of these studies assume at some point in their language or calculations the notion of "identity" or "sameness", something that cannot exist in the corporeal world. This is also why we have disregarded in our considerations the difference drawn in experimental science between "repeatibility" and "reproducibility."

[37] It may be interesting to observe that "statistical mechanics" would not have any sense without such approximations, for the concept of "statistical ensemble" entirely depends thereon.

Strong allowances are made for many of these "idealizations." For instance, the idea of founding science more or less on the notion of repetition (in the approximate sense described before) creates an illusion, the one consisting in thinking that the accumulation of a large number of facts can be used by itself as a "proof" of a theory. Nevertheless, it is evident that facts of the "same kind" (meaning that they are comparable only in certain aspects) are always indefinite in multitude[39] (note, by the way, that we do not say "infinite", see § III.2), so that they can never all be taken into account, quite apart from the consideration that the same facts usually fit several different theories equally well. It will be said that the establishment of a greater number of facts does at least give more "probability" to the theory;[40] but to say so is to admit that any certitude cannot be reached in that way, and that, therefore, the formulated conclusions have nothing "exact" about them. It is also an admission of the totally empirical (and inductive) character of modern science (*cf.* Swann 1940). In that respect, it may be argued that there is nothing better to be done. This is indeed the prevalent position in the modern scientific realm. However, when trying to go beyond the generally established positions one may found some interesting answers. Coomaraswamy (2001, 42) says, for example, that "the *prob*-ability of the relative truths can be established by repeated observations, and such are the statistical 'laws of nature' discovered by science; but behind the experience of order 'there is a further cause of that which is 'always so''; it is because of *eternity* that 'there never was or will be any *time* when movement was not or will not be'; but such a first cause, being itself uncaused, is not *prob*-able but axiomatic (Aristotle, *Phys.* VIII. 1.252 B),—*i.e.*, 'self-revealing', *sva-prakâsa*, 'self-evident'."

These considerations naturally take us to the notion of *statistics*. Statistics consists only in the counting up of a greater or lesser number of facts (events) that are supposed to be exactly alike (if they were not so their addition would be meaningless). It is evident that the picture thus obtained represents a deformation of "reality" (*cf.* note 38), and the less the facts taken into account are alike or really comparable, or the greater the relative importance and complexity of the qualitative elements involved (as in biology, sociology or psychology), the worse is the deformation. Spreading out numbers and calculations gives a kind of illusion of "exactitude"; but, in fact, without this being noticed and because of the strength of preconceived ideas, very divergent conclusions are drawn indifferently from such numbers, so completely without significance are they *in themselves*. The proof of this is that the same statistics in the hands of several experts, even though they all may be "specialists" in the same area, often give rise, according to their respective theories, to quite different conclusions, which may even sometimes be diametrically opposed.[41] In these conditions, the sciences that we call "exact", to the extent that they make use of statistics and go so far as to extract from them predictions for the future (*relying always on the supposed "identicality" of the facts taken into account, whether they are in the future or the past*!), have, to a great extent, a "conjectural" character, as many scientists have been forced to admit. Currently, the tendency is to change the vocabulary from "exact theories" to "rigorous theories", meaning that such theories are checked against experiments, or that these theories are the result of linking up somehow a series of experimental facts,[42] but we think we have already said enough to hint at the scope of these kinds of theories.

---

[38] The "idealizing technique" and its origins are complex and interesting subjects. It is interesting to see, for example, how the "idealization" as understood nowadays has its roots in the Renaissance. An example is given by McMullin (1985): "When Galileo was faced with complex real-world situations... he shifted the focus to a simpler analog of the original problem, one that lent itself more easily to solution. This might, in turn, then lead to a solution of the original complex problem... Idealization in the 'Galilean style' soon became a defining characteristic of the new science" (254-55). Galileo defended these idealizations by arguing that "the departure from truth is imperceptibly small" (256). But, "A physics that borrows its principles from mathematics is thus inevitably incomplete as *physics*, because it has left aside the qualitative richness of Nature" (249). Here, it might be of interest for some to point out that, in the last citation, mathematics should be understood, in our opinion, in its most commonly known quantitative sense, and not, as understood, for example, by the Pythagoreans, in all its qualitative scope (see, *e.g.*, Guthrie [1978, § IV.D.(2)]; D'Olivet [1991, 197-99]; Reghini [1935, §§ II.2 and VI.3]).

[39] This is the case whenever we want to have a theory with some kind of generality. Otherwise, "scientific theories" become nothing else but models describing a *given* number of particular facts.

[40] In this respect, a vivid example is given by Reichenbach (1938, 32) when talking about his book "Wahrscheinlichkeitslehre" (Leiden: Sijthoff, 1935): "As we never have truth, we use a high weight as its representative. Thus my statement about scientific truth as a special case of probability may be interpreted: instead of truth, we always use a representative which is a special case of probability." And he continues setting the main arguments of his theory, one of which is: "there is no difference, on principle, between an observation proposition and a scientific theory as far as their truth character is concerned; both are claimed to be true in respect to some facts, but can never be strictly verified. There is only a difference of degree."

---

[41] The history of modern science offers plenty of examples illustrating this assertion. Here is a striking one with heavy consequences in "daily life" through the influence on people's mentality. During the 70's some specialists, all taking as a departure point statistical data, took two main (radical) positions about the "climate change": on the one hand, there were those who plead for the "global warming" and others, apparently a minority, for the "global cooling"! For a nice review on this "controversy" see Peterson *et al.* (2008). This is just a significant example with very concrete consequences of which everybody is aware; however, almost any modern scientist knows that this kind of debate is part of daily-working-research life of which solutions or explanations come along with a "general consensus" that very often changes, for one reason or another, at rather small intervals of time. Moreover, from these explanations, those that survive for a long enough time are often declared, explicitly or implicitly, "principles" or "fundamental theories", regardless of how aware people are of the relativeness of such ideas. It is then not surprising to read here and there that, "scientifically", "we *have* a comprehensive theory [in this case the Standard Model]... which we believe provides a *complete and correct* description..." and, in this case only one page latter of the same book!, "*beyond doubt...* there must be a theory beyond the Standard Model, and the Standard Model itself *is only an approximation* (albeit a very good one) to the true theory" (Coughlan et al. 2006, ix-x; italics are ours). Here some other significant examples. A renowned French biologist, when talking about the dinosaur's extinction, says that "reconstruire une situation datant de 100 millions d'années est une tentative extrêmement difficile, peut-être chimérique. Tout de nos interprétations est discutable. [reconstructing a situation dating back 100 million years is an extremely difficult endeavour, perhaps even fanciful. Everything about our interpretations is questionable.]" And after setting out a recent hypothesis about the "*Tyrannosaurus*", he adds: "Tout cela en attendant une autre interprétation. [All that while awaiting another interpretation.]" (Grassé 1973, 235-36). In the realm of modern observational cosmology, we have, for example, the abrupt reduction of the value of Hubble's constant from about 600km/s/Mpc in the beginning of 1940s to around 50-100 km/s/Mpc in 1958 (see Peebles 1993). On this matter, it is interesting to quote the opinion of H. Bondi who was considered an authority in the field at the time: "The dominating feature of recent observational work has undoubtedly been the revision of the distance scale, and with it of Hubble's constant, by Baade and Sandage. It is not easy to appreciate now the extent to which for more than fifteen years all work in cosmology was affected and indeed oppressed by the short value of $T$ ($1.8 \times 10^9$ years) so confidently claimed to have been established observationally" (see, *e.g.*, Peebles 1993, 107). It is on these kinds of claims and observational data that modern science relies on. In this context, even the most universal concepts change with an ever increasing speed: for example, in the 1970s the conception of the universe was not at all anymore what astronomers used to believe during former years, that is to say, in conformity with some simple equations of relativity. The discovery of "quasars" was at that moment "shaking" the astronomic community. Astronomers needed to quit the models the formulation of which took so much time. It turned out that the theory of relativity was a poor tool for describing it (see, *e.g.,* Science & Vie nº 691, April 1975**).** About 20 years later, it was completely unexpected to conclude, based on observations of a kind of supernovae and, again, in contradiction with the theories of modern cosmology, that the universe expands... with an acceleration, giving rise once more to a new hypothesis: the well know *dark energy* (see, *e.g.*, Jones & Lambourne 2003).

[42] Another use of the term "rigorous theory", which we shall not treat here, is the one stating that this theory is "mathematically consistent" or, simply, that it relies on some kind of mathematics. (For some interesting considerations on the establishment of the so called "pure science" and its relations with concrete "particular facts" see, *e.g.*, Alston 1971.)

The considerations brought forth so far will help us to study more easily some aspects and conceptions about time in modern science.

### IV.3  Some conceptual positions about time in modern science

**Empty time.** The idea of an empty time, that is, a time without events (or even periods of time without events) is much less widespread than that of empty space (or holed space). This may be the case because time and events are more closely related than are space and bodies: events cannot be moved from one time to another (as bodies do through space by the action of movement). In any case, an empty time has no more effective existence than has an empty space, and in this connection everything that has been said about space could be repeated in some sense: time is coextensive, in the temporal sense, to events, just as space is with bodies; and realized time contains all events, just as realized space contains all bodies. Here, we find, therefore, another symmetry between space and time.

**Homogeneous time.** A homogeneous time can be conceived as something that unrolls itself uniformly. However, as we have already said, this cannot be the case, since quantitatively equal portions of time are filled by totally or partially different sequences of events (*cf.* Granet 1968, 73-80; Coomaraswamy 2001, 17).

**Geometric representation of time.** The strong tendency of looking for a geometrical (or "spatial") representation of time may come from the difficulty—mentioned above—of separating time and space at a sensible level: in order to make measurements of durations one has to relate time back to space. In this respect, it is important to remark that there is no need of any particular geometrical representation for any of the considerations made so far.[43] In modern physics there is a strong practice for representing time geometrically by a straight line. This representation comes from an over-simplification and from the idea that time is something which unrolls itself uniformly (a homogeneous time). This representation may also come from an attempt to describe the unique situation of events in time. However, a straight line is not the only possibility (think, for example, about a helicoidal representation). Some more sophisticated representations of time have been considered with the advent of "space-time" in modern physics (see § V for more details).

**Continuity of time.** A crucial consequence of considering a "discrete time" is the existence of "intervals" devoid of the temporal condition, which we have argued to be impossible within the corporeal world. Indeed, this supposition would imply the hypothesis of a constantly renovated "creation"; otherwise, the world would always vanish away every instant during the intervals of temporal discontinuity.[44] This strongly argues for the extension property of time[45]: it must be extended, just as space is. One could then again develop all that has been said about the divisibility of extension (extension, being continuous, has the inherent quality of divisibility), namely, there cannot be indivisible time durations or smallest time durations nor "final" time durations. Instants conceived of as indivisible are no more parts of duration than are points of space and "final" durations are contrary to the very nature of indefinitude (*cf.* § III.2). In accordance with the same ideas, events (as the analogs of bodies) participate from the duration of time (a one dimensional continuum). There cannot then exist temporally indivisible events. Moreover, movement itself would not be possible if either space or time were discontinuous. Movement is manifested as a combination of space and time[46] and this sort of combination would not be possible if either one (time or space) were discontinuous and the other continuous. Conversely, if both space and time are continuous, whence we can conclude that the movement must be continuous, and, in certain sense, "doubly continuous" (for more developments on this argument see, *e.g.*, Louet [2002, § 3]).

---

[43] In relation to this, it would be interesting to study the representation (not necessarily geometrical) of time by cycles, which has been developed in some ancient doctrines. This conception takes into account a cycle as representing a developed "process", taken in its most general sense: it can be cosmological, historical or other (see, *e.g.*, Coomarswamy 2001, 46; Granet [1988, 80, 85 & 91]; Narayan 2007). Plato put it in this way (*Timaeus* 38b): "ces accidents [que le devenir implique dans l'ordre sensible] sont des variétés du Temps, lequel imite l'éternité et se déroule en cercle suivant le Nombre" (Rivaud 1956, 151) [Those accidents (that the becoming implies in the sensible order) are types of the Time, that imitates eternity and unrolls in a circle following the Number]. A modern western attempt to take this point of view into account, but grounded on the inductive historical method, can be found in the so called "historionomy" or "science of history's laws" founded at the end of the 19th century and beginning of the 20th century (some names going in this direction are Friedrich von Stromer-Reichenbach, Arnold J. Toynbee and François Mentré).

[44] In this respect see also § V.1. However, to fully understand this assertion, we should develop similar considerations about the "now" as those found in Coomaraswamy (2001, 1-7 and 15-16); but this is out of the scope of the present study (see also the Timeo by Plato, in particular 38b).

These considerations strongly argue against the usual conceptions of "time's atomicity." In terms of Kragh *et al.* (1994, 438) "any time interval consists of a finite number of indivisible, but extended and equal parts; the parts may also be unextended so long as they are equally spaced by a finite amount." Modern physicists have developed some considerations (mainly influenced by quantum physics) about the "atomicity or discontinuity of time"; but, as it is well summarized by the same authors,[47] they (the physicists) are not interested in whether "atomistic time" is logically consistent or philosophically meaningful, but whether or not it is of any *use* as an instrument of physics (of modern theories of physics, we would specify). All this is closely related, because of the notion of "space-time", with the considerations developed in § III.3 (in particular note 22).

## V.    Space-time and the "Disappearance" of Time and Space in Leading Physics Theories

In the present state of modern physics, it has become almost inevitable to talk about the concept of "space-time." However, in the limits of the present article, we cannot envisage developing this subject much more than giving some direct and indirect indications. A complete study would be needed to deal with the various aspects around this subject, at least insofar as one's aim is to study it in the same fashion as the present considerations (indeed, since many aspects of "space-time" are "observer dependent", such study should be performed, for example, from the perspective of the *Nyâya*'s point of view of the Hindu doctrine; recall note 5).

The fusion of space and time in a unique entity called "space-time" is usually represented in modern physics as a four-dimensional "geometrical space" (or manifold) with a more or less complicated structure, the coordinates of which get entangled by a common transformation—from one observer to another— law.[48] In this context, time becomes one more geometrical variable, differentiated from the "space" variables by some particular geometric properties.[49] This new structure is chosen in order to respect "causality", "the invariance of light speed", and the "local" equivalence of "physic laws" for any observer (see, *e. g.*, Pathria 2003). In any case, we believe that this tendency of spatializing or geometrizing time comes from the relationship, at the quantitative and sensible level, of space and time: when measuring a time duration, one has to relate it back to space (see § IV.1). From this requirement and the transformation law relating space-and-time coordinates in the theory of relativity, it seems quite "natural" to get to the idea of a "space-time geometry" and things of the sort. The latter represents in addition, seems to us, another example of how modern physics turns to conventional terminology and "idealizations", giving the impression that modern physics theories, to a great extent, are incapable of going beyond a "representation" or description of corporeal nature, of more or less outward appearances (*cf.* note 38). In this sense, it may also be observed that such theories, when going in the direction of a reduction to the quantitative, generally result at the end of the day in a "atomistic" description, that is to say, that these theories introduce discontinuity into their notion of "matter" in such a way as to bring it into much closer relation to the nature of number than that of extension. A clear example is given by the most recent theories of "space-time" where space and time are said to be discrete (see § III.3 and, *e.g.*, Butterfield *et al.* 2000; Monk 1997; Rovelli 2007; 2004; 2006). Are not these approaches going in the direction of bringing everything down to quantity instead of quality, thus moving away from the "essence" of things? Since some of these theories make so many unrealizable suppositions (or "idealizations"), to what extent are they really describing the world we live in? And therefore, has science, as it is conceived presently, any chance of offering somewhat deep explanations of what the world is? In other words, Urban (1924, 2) says that modern science "wishes to know, but not to understand; it seeks knowledge but not intelligibility... The need to represent nature as intelligible may have been abandoned by science."

---

[45] Aristotle seems to arrive at the same conclusion (*Physics*, 220ᵃ 24-220ᵃ 26): "Time is the number of movement in respect of the before and after; and, being the number of what is continuous, it is itself continuous" (see, *e.g.*, Ross 1998, 387). For completeness (*Physics*, 219ᵇ 2-219ᵇ 8): "'number' may mean either that which is numbered, or that by which we number it; time is number in the former sense" (see, *e.g.*, Ross 1998, 386). More developments in this sense can be found in Coomaraswamy (2001, chapter III).

[46] By saying that "movement is manifested as a combination of space and time" we mean that space and time are necessary conditions in order for movement be manifested in the corporeal world. This does not mean, however, that space and time are the first causes of movement, as indicated in § IV.1. We expect to be able to develop this subject in the near future.

[47] Their historical review offers a good overview of physicists' concerns about time "discreteness." Many of the considerations there presented have not lost their topicality.

[48] For further details, we refer the reader to, *e.g.*, Schrödinger (1950), Einstein (1964), Pauli (1967), Misner et al. (1973) and Rovelli (2004).

However that may be, from such conceptions, it is not surprising to hear and read that space and time are "invisible" or "abstract" objects, that they are then thrown away and that the *exclusive* existence of "material" objects and their relative motions is therefore averred. This seems to be the conceptual path followed by the so called "timeless" physics theories, meaning that one manages to get rid of a parameter which was, somewhat conventionally, called "time" or, more generally, to get rid of a (mathematical) structure called "space-time." Concretely, in the context of general relativity[50], the idea is to circumvent the ontological problem of space and time by, roughly speaking, postulating the *identity* between space-time and the gravitational field (or its configurations). Doing so, such approaches try to bring along the idea that, as spelled out by Misner *et al.* (1973, 5), "space acts on matter, telling it how to move. In turn, matter reacts back on space, telling it how to curve." Put in this way, space and time (or space-time) in general relativity are mixed up with the gravitational field and "matter" and it is therefrom concluded that the concepts of space and time are no longer needed. All that is needed to "understand" the (corporeal) world are the gravitational field and "matter." Thus, the "mystery" of space and time is reduced to a semantic (conventional) problem. Following this logic, phrases such as "the end of space and time" or "the non-existence of space and time" at a "fundamental" level are not very surprising.

In our view, the problem in all of this is that, whatever the "beauty" or the potential descriptiveness power of such theories may be, such conclusions are drawn for the corporeal world as a manifested and developing whole, within which—using a rather technical language—"particles" and their interactions coexist (no matter if their are considered "classical" or "quantum"), including the gravitational field (conceived of as the interaction between massive objects). In this sense, space and/or time cannot simply either be thrown away under pretext of being "ad hoc abstract concepts" or be identified with the gravitational field. Space and time are, instead, conditions for the manifestation of the corporeal world, that is to say, these conditions let the corporeal possibilities be developed or, in a rather coarse image, it is what makes bodies to become extended and what makes events succeed one another in the way they do (see also § IV.1). This does not mean that there could not be any kind of relation between these conditions and the things to which they provide a support for their manifestation: bodies and events. In this respect, we are not discarding *a priori* the results obtained by the theory of relativity in all its extension, and for the moment, we content ourselves by saying that we do not see any contradiction between the point of view here proposed and the *facts* confirmed or discovered by the theories giving rise to the space-time concept. Of particular interest would be to study more deeply the relation between space and time and how bodies and events influence space and time, topics expressed in the theory of relativity mainly in terms of the gravitational field, more or less complicated geometries or as transformation of coordinates (in the mathematical sense). On this matter, is the geometrization of time really necessary to understand the nature of time? How far can we push it ahead without moving away from the essence of time? As Skulsky (1938, 54) says, "...having achieved this measure of success in spatializing time, the geometric imagination seeks further satisfaction in insisting that the end of the tale has been reached, that nothing further can or need be said. Time is simply the kind of dimension which appears in the relativistic equation for the spatio-temporal interval, the only difference being the addition of the negative sign"! Some of the conceptual aspects of the relation between space and time were addressed in § IV, especially with respect to measurement matters which are, by the way, particularly important for the conception of the "space-time" notion.

---

[49] This peculiarity is highlighted by demanding the "*t*-coordinate line" to remain inside the local light cone.

[50] Some recent publications about this are: Rovelli (2004, 2006) and Barbour (2000). Somewhat different approaches to the "disappearance" of space-time can be found in Dieks (2006) and Monk (1997).

Finally, we envisage the possibility that bodies and events are influenced by space and time, because of their qualitative aspects. We think, in particular, of the "speed" with which events unfold on time depending on its situation in time. We mean that one should study the possibility that sequences of events comparable one to another do not occupy quantitatively equal durations and the manner in which "speed of events" behaves along time (as far as we know, this issue has not yet been considered in modern physics, but some indications about it can be found, for example, in the Hindu doctrine [Tilak 1903, chapter XIII]). Some quantitative or mathematical aspects could be also therefrom drawn, but our main interest is the qualitative and conceptual aspects of these issues.

## V.1  Substantivalism and relationalism

In contemporary philosophy of science there is a very active debate between substantivalists and relationalists about space, especially in the context of general relativity and, to a smaller extent, in relativistic quantum mechanics. In this context, the debate is grounded on different points of view as regards the gravitational field. On the one hand, the substantivalist position defends the opinion that general relativity is about the description of a dynamical spacetime; and, on the other hand, relationalists claim that such theory is about the description of the gravitational field and that spacetime is at most a mathematical structure, perhaps with a "theoretical" usefulness (see above). In such a situation we agree with the opinion that the debate is reduced to a matter of personal preference, because "the differences between the gravitational field and the other fields are more accidental than essential" (see, *e.g.*, Rovelli [2004, § 2.4.2]). Indeed, the debate becomes rather a question of "personal taste" when it is placed, as it is done in the general relativistic context, within the corporeal world, for it includes, because of its characteristics, the gravitational field.

On the other hand, the "original" philosophical debate is far more complicated (see, *e.g.*, Rynasiewicz 1996. In addition Huggett *et al.* [2009] offer a rather complete list of references on the subject.). However, if we were asked to phrase a position within the limits of the present article, we would say the following: The point of view here presented is "substantivalist", in the sense that space and time are necessary conditions for the manifestation of the corporeal world; but this should not be confused with the rather "standard" modern substantivalism in which space and time can exist independently of "material things" (*cf.* Earman *et al.* 1987; Hoefer 1996). Space, time and corporeal possibilities are equally contained in the total Possibility exposed in § II. On the other hand, these same possibles are coextensive as regards the manifested corporeal world. In this sense, neither space nor time are absolute. In addition, this perspective implies that there is a co-relation between space and time and the (corporeal) possibilities that have to be therein manifested. The latter position may be seen as a kind of "relationalism". Furthermore, it is clear that, within the present perspective, there is in principle no irreducible obstacle in founding a somewhat "operational" or "instrumental" conception of space and time in terms of relations of "bodies" and "events", as those proposed by, *e.g.*, Rovelli (2004, § 2.3) or Bunge (1968). This is not, however, our concern here (in respect to this conception, interesting considerations from the mathematical point of view can be found in Manders 1982).

With regard to the so called "quantum effects" on space and time[51] which we just skimmed over in their "coarse" (often called "classical") interpretation, we have, for example, concluded that, because of the indefinite nature of space and time, "final" or "ultimate" parts of space and time (thus that of the corporeal world) have at most a "theoretical" and "conventional" character, but not a "fundamental" one (see §§ III.3 & IV.3). However, "real" quantum effects, as measurable corporeal phenomena, could be considered, from the standpoint as here briefly outlined, in the sense that these effects are necessarily a "corporeal possibility", and thus contained in the corporeal world. However that may be, many notions must be cleared up before, in particular, that of the dual, corpuscular and undulatory, nature of corporeal "particles." (Some work is already in progress in this direction.) In this respect, we would like to retain the following reflection which, though written in 1951, seems to us of a striking topicality: "Is the impossibility of continuous, gapless, uninterrupted description in space and time really founded in incontrovertible facts? The current opinion among physicists is that, that this *is* the case. Bohr and Heisenberg have put forward a very ingenious theory about it, which is so easy to explain that it has entered most popular treatises on the subject—I should say, unfortunately; for its philosophical implication is usually misunderstood. I am going to argue against it..." (Schrödinger 1996, 153).

Finally, so far we have so far considered the space *tout court* as the contraction of "realized corporeal space" (recall note 12) without making a great deal of the difference between space as a condition and the "substance" it must contain. Indeed, a careful reflection on the present considerations should bring out the following: space, being only a container, that is to say, a condition of existence and not an independent entity (as some "absolutists" retain), *cannot* as such be *the substantial principle* of bodies. Therefore, "realized space" implies, on the one hand, space as condition of existence, the container; and, on the other hand, the "substance" (the content) which, considered in its principle, *i.e.*, in its state of primordial indifferentiation, has to contain potentially not only all the "elements" (in the Greek sense), but also all bodies. It is only at this fundamental point or limit that one can talk about a "homogeneous space", in the sense that the homogeneity of the substantial principle contained by space renders it suitable to receive all forms in its modifications (it is also at this point, we think, that the Leibniz' "relationalism" has to be sought). However, "homogeneous space" in this sense has, properly speaking, no existence at all, being nothing more than a virtuality out of which "instantaneously" (at an instant, thus "a-temporally", and not in "no-matter-how-short a fraction of time duration") corporeal world is manifested. To fully understand this point, we believe that it may be very helpful to seriously and without prejudices take into account both the notion of "the now" as found in many different doctrines (see, *e.g.*, Coommaraswamy 2001) and the "production of numbers or geometry" as dealt with for example in Pythagorism or Far-East doctrine (see, *e.g.*, D'Olivet [1991, 197-99]; Reghini [1935, §§ II.2 and VI.3]).

---

[51] Some pictures of space-time taking into account also notions of quantum theory can be found, for example, in the review by Monk (1997) and in the articles of Butterfield *et al.* (1999, 2000).

## VI.   Conclusion and Perspectives

Our intention in the present article was to give a clear exposition of the notion of space and time as realized conditions of the corporeal world, respectively, the container of bodies and the container of events. The language used and point of view taken in this exposition is rooted in an uncontradictable or axiomatical basis (the Total possibility) common to many doctrines in the East and the West. In addition we have tried to show how such an approach can be useful to clear up many conceptual and "philosophical" aspects and paradoxes of modern physics. Instead of summarizing such results, we prefer to close this work with some more "applications" and end up with some rather general considerations about the conception of science.

Take as example Kant's famous "cosmological" antinomy: *the world had a beginning in time, and is also limited in regard to space, or the world had no beginning, and has no limits in space, but it is infinite as regards both time and space* (*cf.* Kant 1988, 135). This antinomy can be solved employing the ideas presented in this work, as follows: it is the coextension of space and time with the corporeal world which permits us to assert that there cannot be more space than there is corporeal world (and vice versa, as it is generally accepted); and, similarly, that time began with the corporeal world, and this world with time. On the other hand, neither space nor time (and, therefore, nor the corporeal world) can be "infinite". What we can say at most is that they extend "indefinitely." In other words, corporeal manifestation is no more infinite than is space itself, for, like space, it does not contain every possibility, but only represents a certain particular order of possibilities, and it is limited by the determinations that constitute its very nature (see § II for details). Similarly, the corporeal world is not eternal ("infinite as regards time") because it is contingent, eternity being another "aspect", insofar as it is permissible to say so, of Infinite. Here, eternity should not be confused with the notion of perpetuity, which in general means indefinite as regards time or any other mode of duration (*cf.* § III.2 and note 30).

Observe, in addition, that the principal current "cosmological" observations (referring here only to the corporeal world) are comprised in the point of view here developed. Indeed, most of these results establish that the space of the "actual universe" is expanding and that its observable limits are, in spite of the (amazing) technical development, more and more out of our reach. The main interest today is to determine the reason of the acceleration of this expansion. Currently the explanations are given in terms of the cosmological constant, the "quantum vacuum", the "quintessence", and so forth (see, *e.g.*, Jones & Lambourne 2003). In any case, the indefinite space here presented is consistent with any of these current observations, which, by the way, are reaching their limits, not only for almost any conceivable instrument, but also for present-state-of-art physics. Modern radio telescopes are reaching the cosmic microwave background radiation threshold, beyond which even the most powerful telescopes cannot detect any electromagnetic radiation. Some scientists are thereupon frustrated, others are rather happy because this means "new" physics and, above all, more technical developments, in particular, for gravitational wave detectors.[52] It is as if the latter were pleased by the mere fact of exploring a "universe", the limits of which are being constrained, to declare it as unreachable by any means.[53] Such a conclusion would not surprise us, because, as we have explained, the indefinite is something analytically inexhaustible and modern science, with all its means of knowledge, is essentially analytical. In our opinion, so far as science does not radically change its point of view toward a more "synthetic" perspective, this will be inevitably the conclusion for the majority.

---

[52] An example is given by the NASA/ESA project that plans to launch by 2016 an instrument to directly measure gravitational waves (LISA), see, *e.g.*, http://lisa.nasa.gov/technology/.

In this respect, we believe that the point of view here offered presents many advantages, the main interest of which is to take into account the qualitative aspects of the object under study, of course, without disregarding the quantitative aspects when they exist. In particular, we have studied some aspects of space and time that are closer to quality than to quantity. Modern science, on the contrary, takes almost exclusively the empirical and practical perspective, plus an analytic method. Its point of departure is mainly the mechanistic and materialistic conceptions, together with an almost exclusively quantitative point of view, entailing necessarily a lack of interest in the deep sense of things, *i.e.*, in their principles or in "synthetic" knowledge (we recall what we have said in notes 4 & 15). In this perspective, it is not surprising to see that most modern scientists content themselves by measuring some properties (those with a more direct quantitative character) of the object under study, or by finding a possible "practical" application for them, disregarding in this way the meaning of the concepts they are using. As a natural consequence, theoretical science is an ever-changing[54] field out of which it is quite difficult to get a general and coherent picture. Scientific theories, being almost exclusively attached to experimental facts, to the ever-changing sensible world, lack true principles.

In this sense, we have illustrated, through some examples, how modern physics turns to conventional terminology and "idealizations" that are found to be unable, mainly because of their analytical character, to penetrate the "representation", or description it makes, of the corporeal nature. This causes the clear impression that physics theories, to a great extent, are not much more than a "representation" or description of corporeal nature, of outward appearances, lacking to that same extent of a truly explanatory value. Furthermore, all scientific approaches seem to point in the direction of bringing everything down to quantity instead of quality, thus moving away from the "essence" of things. Clearly, we do not completely discredit the usefulness (mainly in "practical life" and technical development) or the descriptive capacity of some modern scientific theories. Nevertheless we sincerely wonder to what extent, because some of these theories make so many unrealizable suppositions (or "idealizations"), they are really describing the world we live in and, therefore, if science, as it is conceived presently, has any chance of offering somewhat deep explanations? (The deepest explanations should be, because of the analytic character of science, already discarded.)

Our scientific position differs radically from that of some physicists, like G. Kirchhof and E. Mach, who, under the influence of D. Hume's conclusion that the relation between cause and effects is not directly observable and enunciates nothing but the regular succession, maintained that natural science cannot warrant any explanations, that it aims only at, and is unable to attain anything but, a complete and economical description of the observed facts (*cf.* § IV.2). We believe that science, when linked up to "superior or first principles", can overcome its "natural" limitations.[55] Science grounded on the "immutable" principles finds its *raison d'être*. We hope that the present article helps to give an idea on how this has been done elsewhere in the great doctrines expressed by humanity. These considerations are based in "Natural Philosophy" or, better, in "Cosmology", in the sense that the scope of the cosmological order includes the corporeal world, and, in general, the "Manifested World" (see § I). This constitutes part of science in its extended sense, *i.e.*, science supported by principles (§ II), in other words, scientific knowledge supported by metaphysical knowledge or reasoning supported in, using the Aristotelian terminology, intuitive reason (or intellect). In this sense, science is rational, discursive knowledge, always indirect, a knowledge by reflection; metaphysics is instead an intuitive and unmediated knowledge (this intellectual intuition must not be confused with the intuition spoken by certain philosophers, which is of sensible order).

---

[53] In modern cosmology, for example, if the inflatory hypothesis of A. Guth is verified, the "observable universe" would be just a "tiny" part of the "entire universe."

[54] The historical account written by Redhead (1980) offers an example in contemporary physics of this statement.

[55] This is, for example, what is done in Hindu doctrine when the *Vaishêshika* point of view is linked up with the *Sânkhya* one, *i.e.*, "Cosmology" as regards its principles (recall note 5). It seems that some Greek "schools" and some Scholastics of the Middle Age included this point of view in their "Cosmology", but it was less clearly distinguished.

After all, Kant himself admits the limited nature of reason (therefore of science) when he claims that the greatest, and perhaps the only, use of all philosophy of pure reason is, accordingly, of a purely negative character, since it is not an instrument for extending knowledge, but a discipline for limiting it (*cf.* Kant 1988, 234). Aristotle explains how this limitation can possibly be "circumvented" (*Posterior Analytics, 100$^b$5-100$^b$ 17*): "... no [thinking] state is superior to science except intuitive reason; the first principles are more knowable than the conclusions from them, and all science involves the drawing of conclusions. It follows that it is not science that grasps the first principles; and that it must be intuitive reason that does so. This follows also from the fact that demonstration cannot be the source of demonstration, and therefore science cannot be the source of science; if, then, intuitive reason is the only necessarily true state other than science, it must be the source of science. It apprehends the first principles, and science as a whole grasps the whole subject of study; *i.e.* science as a whole grasps its object with the same certainty with which intuitive reason grasps the first principles" (see Ross [2001, 675 and 678]). St. Thomas of Aquin confirms (*De Veritate*, q. XV, a. 1) by saying: "*Ratio discursum quemdam designat, quo ex uno in aliud cognoscendum anima humana pervenit; intellectus vero simplicem et absolutam cognitionem (sine aliquo motu vel discursu, statim in prima et subita acceptione) designare videtur*" (See, *e.g.*, Aquinatis 1953, 307). [The reason refers to a given discursive process through which the human soul reaches the knowledge of one thing by using another one; the intellect refers to the simple and absolute truth's knowledge (which is obtained immediately, without any movement nor any discursive process, in a first and sudden grasping).] Finally, Eastern doctrines accomplish: "The mind, with its thoughts perceived by the Self, is *drishya* ('objet'), and the Self is *drik* ('sujet')" (Maharashi 1997, 174). Moreover, "la connaissance du Soi ne dépend pas d'une preuve (autre que lui-même), car seule la détermination d'une chose contestable dépend d'une preuve et le Soi n'est pas dans ce cas. La démonstration que la réalité du Soi dépend d'une preuve dépendrait elle-même du Soi. Ainsi il appert que celui qui ferait usage de cette preuve est lui-même le Soi. [the knowledge of the Self does not depend on any proof (besides Itself), for only the establishing of something contestable depends on a proof, while the Self is not in this case. The very demonstration that the reality of the Self depends on a proof would depend on the Self. It follows therefore that who would use this proof is the Self.]" (Shankarâchârya 2003, 32). Finally, in words of the Taoist Lie-tzeu: "Uni au Principe, il connaît tout par les raisons générales supérieures, et n'use plus, par suite, de ses divers sens, pour connaître en particulier et en détail. [Joined to the Principle, he knows everything through the general superior reasons, and he, therefore, no longer uses anymore his different senses to know in particular and in detail.]" (Weiger 1950, 129).[56]

We would definitely like to close this article with the following considerations. Many readers will undoubtedly wonder how these considerations have any relation with "real science." We would like to answer in the words of one of the founders of quantum mechanics: "Science aims at nothing but making true and adequate statements about its object. The scientist only imposes two things, namely truth and sincerity, imposes them upon himself and upon other scientists" (Schrödinger 2010, 117). "[T]he isolated knowledge obtained by a group of specialists in a narrow field has in itself no value whatsoever, but only in its synthesis with all the rest of knowledge and only inasmuch as it really contributes in this synthesis something toward answering the demand τίνες δὲ ἡμεῖς ('who are we?'). [...] specialization is not a virtue but an unavoidable evil... all specialized research has real value only in the context of the integrated totality of knowledge" (Schrödinger 1996, 109-111). Elsewhere he continues: "It is pathetically amusing to observe how on the one side [modern science] only scientific information is taken seriously, while the other side [metaphysics[57]] ranges science among man's worldly activities, whose findings are less momentous and have, as a matter of course, to give way when at variance with the superior insight gained in a different fashion, by pure thought or by revelation. One regrets to see mankind strive towards the same goal along two different and difficult winding paths, with blinkers and separating walls, and with little attempt to join all forces and to achieve, if not a full understanding of nature and the human situation, at least the soothing recognition of the intrinsic unity of our search. This is regrettable, I say, and would be a sad spectacle anyhow, because it obviously reduces the range of what could be attained if all the thinking power at our disposal were pooled without bias. [...] As we scan its windings over hills and vales back in history we behold a land far, far, away at a space of over two thousand years back, where the wall flattens and disappears and the path was not yet split, but was only *one*. [...] the true subject was essentially one, and that important conclusions reached about any part of it could, and as rule would, bear on almost every other part. The idea of delimitation in water-tight compartments had not yet sprung up" (Schrödinger 1996, 11-14). "In fact, if we cut out all metaphysics it will be found vastly more difficult, indeed probably quite impossible, to give any intelligible account of even the most circumscribed area of specialisation within any specialised science you please. Metaphysics includes, amongst other things—to take just one quite crude example— the unquestioning acceptance of a more-than-physical—that is, transcendental—significance in a large number of thin sheets of wood-pulp covered with black marks such as are now before you" (Schrödinger 1964, 3). Indeed, E. Schrödinger had tried, it seems to us, to sincerely and prudently put no limits to his scientific questioning; even inquiring, with some satisfactory results, in the "Vedantic vision"... (*cf.* Schrödinger 1964). In this respect, as Urban (1938, 281-84) put it, "if science is possible *only* if metaphysics is possible" and if "a large part of scientific knowledge *is* symbolic", then with the recognition of these facts must come a change in our conceptions of scientific knowledge. Furthermore, perhaps the symbolic character of science is the way out for the impasse in which modern science seems to be... This is in any case the proposal of most doctrines from the West and the East for overcoming the natural limits of any science.

---

[56] The list of similar quotes, the coherence of which is to be noticed, could be increased very easily. An example in Islamic doctrine can be found, for example, in Asín (1941, 70).

## Acknowledgments

We thank R. G. and our Italian friends without whom this work would have not been at all possible. We are also grateful to Ben for English corrections. M. M. thanks the financial support provided during his PhD at *Centre de Physique Théorique* (France) by the Programme Alßan, the European Union Programme of High Level Scholarships for Latin America, scholarship No. E04D033873MX and by the Programme SFERE-CONACYT (France-Mexico), fellowship No. 166826, which partly made possible this work.

---

[57] Here, by "metaphysics" the author means whatever domain or knowledge out of the reach of physics in its usual conception (*i.e.*, the study of the corporeal world) and not, as we had the intention to mean before, that which is beyond the "manifested world" or, in more "effective" terms, that which is beyond the "natural philosophy" in all its conceivable extension, *i.e.*, the knowledge or grasping, insofar as this expression is permitted, of Principles (recall notes 5 & 6).